\documentclass[english,british]{article}
\usepackage[T1]{fontenc}
\usepackage[latin9]{inputenc}
\usepackage[a4paper]{geometry}
\usepackage{color}
\usepackage{verbatim}
\usepackage{textcomp}
\usepackage{amstext}
\usepackage{graphicx}
\usepackage{amssymb}

\makeatletter

\newcommand{\lyxdot}{.}

\usepackage[dvips] {epsfig}
\usepackage{color}
\usepackage{graphicx}

\makeatother

\usepackage{babel}

\begin{document}

\title{Diversity Spectra of Spatial Multipath Fading Processes  }

\author{Henrik ~Schulze \\
South Westphalia University of Applied Sciences, \\
Meschede, Germany}

\maketitle
%
{}
\begin{abstract}
We analyse the spatial diversity of a multipath fading process for
a finite region or curve in the plane. By means of the Karhunen-Loève
(KL) expansion, this diversity can be characterised by the eigenvalue
spectrum of the spatial autocorrelation kernel. This justifies to
use the term \emph{diversity spectrum} for it. We show how the diversity
spectrum can be calculated for any such geometrical object and any
fading statistics represented by the power azimuth spectrum (PAS).
We give rigorous estimates for the accuracy of the numerically calculated
eigenvalues. The numerically calculated diversity spectra provide
useful hints for the optimisation of the geometry of an antenna array.
Furthermore, for a channel coded system, they allow to evaluate the
time interleaving depth that is necessary to exploit the diversity
gain of the code.

\end{abstract}

\section{Introduction }

Multipath fading is a major source of degradations in wireless digital
communication systems \cite{Proakis,BenedettoBiglieri,Rappaport}.
Information bits that fall into deep fades may get lost if no measures
are taken against this. All methods to prevent this loss of information
employ some kind of diversity, i.e. they spread the information over
a wider range of the channel so that enough reliable parts of it
can be exploited by the transmission system. In its simplest form,
diversity is employed by multiple reception of the same information
bits at different frequencies, time slots, or antenna locations. More
 efficient ways to employ diversity is channel coding in conjunction
with an appropriate time and frequency interleaving \cite{Proakis,BenedettoBiglieri,Rappaport,SchulzeLueders}
as applied in several systems. Multiple-input-multiple-output (MIMO)
systems with several transmit and receive antennas are an efficient
method to increase the channel capacity \cite{Foschini1998,Telatar1999,KuehnBook}.
To achieve the maximal capacity, these antennas must be uncorrelated,
which requires a sufficient spatial distance between them. 

In practice however, the information can only be spread over a finite
region of the channel. Therefore, there are channel correlations that
diminish the possible gains of all of these methods. In this paper
we provide tools to quantify these degradations. We concentrate on
space and time diversity aspects of the multipath channel and leave
aside frequency diversity as a topic of its own. Space and time diversity,
however, are closely related for the scenario of a receiver that is
moving through a spatial fading pattern: If the receive antenna moves
along a straight line, the time fading amplitude as a function of
time is nothing but the fading amplitude as a function of the space
variable, scaled by the velocity of the vehicle. Thus, time correlations
are nothing but scaled space correlations. The diversity of such
a given space or time interval can be quantified by evaluating the
equivalent uncorrelated diversity branches of the fading process.
For a time interval $\left[t_{1},t_{2}\right]$, these uncorrelated
branches can be obtained by the Karhunen-Loève (KL) expansion \cite{Van_Trees,Davenport_Root,Papoulis}
of the fading process $a\left(t\right)$ given by\begin{equation}
a\left(t\right)=\sum_{n=1}^{\infty}c_{n}\varphi_{n}\left(t\right)\,,\mbox{ }t\in\left[t_{1},t_{2}\right]\,.\label{eq:KLE-1}\end{equation}
In that expansion, the basis functions $\varphi_{n}\left(t\right)$
are orthogonal and of normalised energy, and the coefficients $c_{n}$
obey the condition\begin{equation}
\mathrm{E}\left\{ c_{m}c_{n}^{\ast}\right\} =\lambda_{n}\delta_{mn}\,.\label{eq:c_n_vs_lambda_n}\end{equation}
This means that the process can be decomposed into a series of orthogonal
and uncorrelated components, each with energy $\lambda_{n}$. We may
thus call them \emph{equivalent uncorrelated diversity branches}.
Their respective strenghts are characterised by the values $\lambda_{n}$,
which are the eigenvalues of an integral equation with kernel given
by the autocorrelation function (ACF) of $a\left(t\right)$ \cite{Van_Trees,Davenport_Root,Papoulis}.
The eigenvalue spectrum $\left\{ \lambda_{n}\right\} _{n=1}^{\infty}$
characterises the amount of diversity gain that can be achieved in
the time interval under consideration. We will thus call $\left\{ \lambda_{n}\right\} _{n=1}^{\infty}$
the \emph{diversity spectrum} of the stochastic process corresponding
to that interval. Coding with interleaving is closely related to diversity.
The error event probabilities corresponding to the Hamming distance
$d$ of a convolutional code with perfect interleaving in a Rayleigh
fading channel are asymptotically proportional to $SNR^{-d}$, i.e.
they behave like the error probabilities of $d$ uncorrelated diversity
branches \cite{Proakis,BenedettoBiglieri,SchulzeLueders}. For interleaving
over a finite time interval, this error probability will decrease
much slower if there are less than $d$ eigenvalues of significant
size \cite{Schulze2010a}. 

In that example where we consider time variance that is caused by
vehicle motion, the diversity spectrum can equivalently and uniquely
assigned to a corresponding spatial interval (the segment of a straigth
line). On the other hand, a spatial interval by itself can be interpreted
as the geometrical base of a linear antenna array. In that case the
term (antenna)\emph{ aperture} is usually applied. With the continuous-limit
idealisation that the straigth line is densely covered with sensors,
the eigenvalue spectrum may then be interpreted as the multipath richness
of the fading process (or the electromagnetic field) in that region.
 We may thus say that the eigenvalue analysis of a time interval
and densely covered uniform linear array (ULA) are equivalent.

Our goal is to calculate the diversity spectrum not only for a given
spatial (or time) interval as discussed above, but for quite general
one- or two-dimensional apertures in the plane. The two-dimensional
aperture of a rectangle, e.g., arises naturally if a linear antenna
array is mounted on a vehicle, perpendicular to the direction of motion.
One edge of the rectangle corresponds to the length of the array and
the perpendicular edge corresponds to the route covered by the vehicle
during a given time interleaving depth.

Rather than characterising the diversity of a given aperture by the
complete diversity spectrum $\left\{ \lambda_{n}\right\} _{n=1}^{\infty}$,
it is desirable to have a single number that reflects the diversity
degree and can be interpreted in some way as the effective number
of equivalent independent diversity branches. It turns out that the
\emph{diversity measure} defined by\begin{equation}
\omega=\frac{\left(\sum_{i=1}^{\infty}\lambda_{i}\right)^{2}}{\sum_{i=1}^{\infty}\lambda_{i}^{2}}\,\label{eq:Diversity-Measure}\end{equation}
is a useful quantity that allows such an interpretation. This definition
is the dense-array limit of the $L$-antenna diversity measure\begin{equation}
\omega=\frac{\left(\sum_{i=1}^{L}\lambda_{i}\right)^{2}}{\sum_{i=1}^{L}\lambda_{i}^{2}}\label{eq:Diversity-Measure-finite}\end{equation}
that has been discussed in-depth by \cite{Ivrlac2003} and \cite{Muharemovic2008}.
For MIMO systems, one can calculate the diversity measures $\omega_{Tx}$
and $\omega_{Rx}$ at the transmitter (Tx) and receiver (Rx) site,
respectively. Under the assumption that there are no correlations
between both sites, the spectral efficiency slope $\Sigma_{0}$ of
the low power regime \cite{Lozano2003,Muharemovic2008} is given by
\begin{equation}
\frac{2}{\Sigma_{0}}=\frac{1}{\omega_{Rx}}+\frac{1}{\omega_{Tx}}\label{eq:Slope-Formel}\end{equation}
Our method allows to calculate $\Sigma_{0}$ for dense arrays. 

The major part of our work is devoted to the solution of the eigenvalue
problem in the Hilbert space of square integrable functions over a
given aperture. In \cite{Kennedy2007}, this problem has been solved
for the special case where the aperture is a disk of finite radius.
The authors approximate the fading process by a truncated orthogonal
series expansion based on Bessel functions, and they prove truncation
error bounds for the fading process. Based on that rigorous bounds,
they heuristically argue that a good approximation of the process
itself leads to a good approximation of the eigenvalues $\lambda_{n}$.

We adopt the method of \cite{Kennedy2007} and put it into a more
general Hilbert space framework. This allows us to calculate diversity
spectra not only for a disk, but for quite general curves or areas
in the two-dimensional plane. Furthermore, by utilising the theory
of compact operators, we are able to prove rigorous bounds of the
approximation errors for the eigenvalues themselves. We give several
examples of numerically calculated eigenvalue spectra and diversity
measure for scenarios of several apertures and scattering environments
with azimuth spectra (PAS) of different degrees of anisotropy.

The rest of the paper is organised as follows. In Section 2, we introduce
the channel model and discuss its statistical properties. In Section
3, we introduce the Hilbert space of square integrable functions on
the aperture and define the the autocorrelation kernel as an operator
acting on that Hilbert space. Numerical examples of diversity spectra
and diversity measures for different scenarios are presented in Section
4. In Section 5, we summarise our results and draw some conclusions.
The necessary Hilbert space framework of our work as well as the proofs
of the theorems can be found in the Appendix.

\section{The Statistical Model for Spatial Multipath Fading}

\subsection{Solutions of the Helmholtz Equation}

We consider scalar spatial multipath interference (\emph{fading})
patterns that are given as scalar solutions $\phi$ of the two-dimensional
(2D) Helmholtz equation  \begin{equation}
\vartriangle\phi+k^{2}\phi=0\,,\label{eq:Helmholtzgleichung}\end{equation}
where $\vartriangle=\partial^{2}/\partial x^{2}+\partial^{2}/\partial y^{2}$
is the 2D Laplacian operator and $k$ is the wave number. The Helmholtz
equation (HE) is equivalent to the homogeneous wave equation for a
fixed wavelength. Considering only the homogeneous equation means
that all scatterers are outside the region of interest. Since we keep
the wavelength fixed, we may set it equal to one which means that
we measure the space variable $\mathbf{x}\in\mathbb{R}^{2}$ in units
of wavelengths. With that normalisation, the wave number is simply
given by $k=2\pi$. 

The prototype solution of the HE is given by \begin{equation}
\phi\left(\mathbf{x}\right)=\mathrm{e}^{j\mathbf{k}\cdot\mathbf{x}}\,,\label{eq:spezielle_Lsg}\end{equation}
with $\left|\mathbf{k}\right|=k=2\pi$. The wave vector $\mathbf{-k}$
points into the direction of propagation, thus $\mathbf{k}$ points
into the direction of the incoming wave. We write $\mathbf{k}$ and
$\mathbf{x}$ in polar coordinates as \begin{equation}
\mathbf{k}=2\pi\left(\begin{array}{c}
\cos\alpha\\
\sin\alpha\end{array}\right)\,\mbox{\,\ and }\,\,\mathbf{x}=r\left(\begin{array}{c}
\cos\beta\\
\sin\beta\end{array}\right)\,,\label{eq:k_und_x}\end{equation}
note that\begin{equation}
\mathbf{k}\cdot\mathbf{x}=2\pi r\cos\left(\alpha-\beta\right)\label{eq:-kx-Skalarprodukt}\end{equation}
and obtain \begin{equation}
\phi\left(\mathbf{x}\right)=\mathrm{e}^{j2\pi r\cos\left(\alpha-\beta\right)}\,.\label{eq:spezielle_Lsg_polar}\end{equation}
The angle $\alpha$ is the \emph{angle of arrival} (AoA). 

A general solution $a\left(\mathbf{x}\right)$ of the HE can be obtained
as a superposition of solutions of type (\ref{eq:spezielle_Lsg_polar})
with different angles of arrival $\alpha$:\begin{equation}
a\left(\mathbf{x}\right)=\int_{-\pi}^{\pi}\hat{a}\left(\alpha\right)\mathrm{e}^{j2\pi r\cos\left(\alpha-\beta\right)}\mathrm{d}\alpha\label{eq:a(x)}\end{equation}
In that equation, $\hat{a}\left(\alpha\right)$ is a weighting factor,
and $\hat{a}\left(\alpha\right)\mathrm{d}\alpha$ has to be interpreted
as the part of the fading amplitude $a\left(\mathbf{x}\right)$ corresponding
to AoAs corresponding to the directions inside the interval $\left[\alpha,\alpha+\mathrm{d}\alpha\right]$.
We expand the weighting factor $\hat{a}\left(\alpha\right)$ into
a Fourier series\begin{equation}
\hat{a}\left(\alpha\right)=\sum_{n=-\infty}^{\infty}\hat{a}_{n}\mathrm{e}^{j\alpha n}\label{eq:FR_a_dach}\end{equation}
with coefficents\begin{equation}
\hat{a}_{n}=\int_{-\pi}^{\pi}\mathrm{e}^{-j\alpha n}\hat{a}\left(\alpha\right)\frac{\mathrm{d}\alpha}{2\pi}\,.\label{eq:FK_a_dach}\end{equation}
For convenience, we define\begin{equation}
\tilde{a}_{n}=2\pi\hat{a}_{n}=\int_{-\pi}^{\pi}\mathrm{e}^{-j\alpha n}\hat{a}\left(\alpha\right)\mathrm{d}\alpha\,.\label{eq:def-a_n-tilde}\end{equation}
Inserting the Fourier series (\ref{eq:FR_a_dach}) into Equation (\ref{eq:a(x)})
we obtain: \begin{eqnarray*}
a\left(\mathbf{x}\right) & = & \int_{-\pi}^{\pi}\hat{a}\left(\alpha\right)\mathrm{e}^{j2\pi r\cos\left(\alpha-\beta\right)}\mathrm{d}\alpha\\
 & = & \sum_{n=-\infty}^{\infty}\hat{a}_{n}\int_{-\pi}^{\pi}\mathrm{e}^{j\alpha n}\mathrm{e}^{j2\pi r\cos\left(\alpha-\beta\right)}\mathrm{d}\alpha\\
 & = & \sum_{n=-\infty}^{\infty}\hat{a}_{n}\int_{-\pi}^{\pi}\mathrm{e}^{j\left(\alpha+\beta\right)n}\mathrm{e}^{j2\pi r\cos\left(\alpha\right)}\mathrm{d}\alpha\\
 & = & \sum_{n=-\infty}^{\infty}\hat{a}_{n}\mathrm{e}^{j\beta n}\int_{-\pi}^{\pi}\mathrm{e}^{j\alpha n}\mathrm{e}^{j2\pi r\cos\left(\alpha\right)}\mathrm{d}\alpha\end{eqnarray*}
The complex version of Bessel's first integral (\cite{ArfkenWeber},
p. 690),\begin{equation}
\mathrm{J}_{n}\left(x\right)=\frac{1}{2\pi j^{n}}\int_{-\pi}^{\pi}\mathrm{e}^{j\alpha n}\mathrm{e}^{jx\cos\left(\alpha\right)}\mathrm{d}\alpha\,,\label{eq:Besselsches_Integral-1}\end{equation}
then leads to the series expansion\begin{equation}
a\left(\mathbf{x}\right)=\sum_{n=-\infty}^{\infty}\tilde{a}_{n}\mathrm{e}^{j\beta n}j^{n}\mathrm{J}_{n}\left(2\pi r\right)\,.\label{eq:a(x)_Bessel-1}\end{equation}

\subsection{Properties of the Fading Random Process}

The fading amplitude $a\left(\mathbf{x}\right)$ is modeled as a zero-mean
stochastic process. This means that $\hat{a}\left(\alpha\right)$
in Equation (\ref{eq:a(x)}) is a family of random variables depending
on the angle $\alpha$. We impose the uncorrelated \emph{scattering
assumption} (US) \cite{Bello1963,Fleury2000} which means that waves
impinging from different directions can be regarded as uncorrelated:\begin{equation}
\mathrm{E}\left\{ \hat{a}\left(\alpha\right)\hat{a}^{\ast}\left(\alpha^{\prime}\right)\right\} =\mathcal{S}\left(\alpha\right)\delta\left(\alpha-\alpha^{\prime}\right)\label{eq:US-Annahme}\end{equation}
The quantity $\mathcal{S}\left(\alpha\right)$ is called the \emph{power
azimuth spectrum} (PAS), and $\mathcal{S}\left(\alpha\right)\mathrm{d}\alpha$
has to be interpreted as the amount of average power that is impinging
from directions inside the interval $\left[\alpha,\alpha+\mathrm{d}\alpha\right]$.
The PAS must be positive and integrable. We normalize the PAS to the
total power one: \begin{equation}
\int_{-\pi}^{\pi}\mathcal{S}\left(\varphi\right)\mathrm{d}\varphi=1\label{eq:Leistungsnormierung}\end{equation}
From Equation (\ref{eq:a(x)}), together with the US assumption (\ref{eq:US-Annahme}),
we obtain the following expression for the autocorrelation of the
fading process:\begin{equation}
\mathrm{E}\left\{ a\left(\mathbf{x}\right)a^{\ast}\left(\mathbf{x}^{\prime}\right)\right\} =\int_{-\pi}^{\pi}\mathrm{e}^{j2\pi r\cos\left(\alpha-\beta\right)}\mathrm{e}^{-j2\pi r^{\prime}\cos\left(\alpha-\beta^{\prime}\right)}\mathcal{S}\left(\alpha\right)\mathrm{d}\alpha\label{eq:ACF-Transl-Invar-1}\end{equation}
Using Equation (\ref{eq:k_und_x}), we note that\[
r\cos\left(\alpha-\beta\right)-r^{\prime}\cos\left(\alpha-\beta^{\prime}\right)=\mathbf{k}\cdot\left(\mathbf{x}-\mathbf{x}^{\prime}\right)\,.\]
Thus, $\mathrm{E}\left\{ a\left(\mathbf{x}\right)a^{\ast}\left(\mathbf{x}^{\prime}\right)\right\} $
depends only on the difference variable, $\mathbf{x}-\mathbf{x}^{\prime}$,
and we may write\begin{equation}
\mathrm{E}\left\{ a\left(\mathbf{x}\right)a^{\ast}\left(\mathbf{x}^{\prime}\right)\right\} =\rho\left(\mathbf{x}-\mathbf{x}^{\prime}\right)\label{eq:AKF-1}\end{equation}
with the spatial autocorrelation function $\rho\left(\mathbf{x}\right)$
given by\begin{equation}
\rho\left(\mathbf{x}\right)=\int_{-\pi}^{\pi}\mathrm{e}^{j2\pi r\cos\left(\alpha-\beta\right)}\mathcal{S}\left(\alpha\right)\mathrm{d}\alpha\,.\label{eq:Def_rho}\end{equation}
From Equation (\ref{eq:Leistungsnormierung}) we obtain $\rho\left(0\right)=\int_{-\pi}^{\pi}\mathcal{S}\left(\varphi\right)\mathrm{d}\varphi=1\,.$
Thus, the autocorrelation function has the property\begin{equation}
\left|\rho\left(\mathbf{x}\right)\right|\le1\,.\label{eq:AKF_le_1}\end{equation}

The PAS has a Fourier expansion\begin{equation}
\mathcal{S}\left(\alpha\right)=\sum_{n=-\infty}^{\infty}\hat{s}_{n}\mathrm{e}^{j\alpha n}\label{eq:FR-PAS}\end{equation}
with coefficients given by\begin{equation}
\hat{s}_{n}=\int_{-\pi}^{\pi}\mathrm{e}^{-j\alpha n}\mathcal{S}\left(\alpha\right)\frac{\mathrm{d}\alpha}{2\pi}\,.\label{eq:s_dach-1}\end{equation}
We assume that $\mathcal{S}\left(\alpha\right)$ is a piecewise continuous,
bounded function over the interval $\left[-\pi,\pi\right]$. Then
the Fourier series (\ref{eq:FR-PAS}) for $\mathcal{S}\left(\alpha\right)$
converges almost everywhere. For convenience, we define\begin{equation}
\tilde{s}_{n}=2\pi\hat{s}_{n}=\int_{-\pi}^{\pi}\mathrm{e}^{-j\alpha n}\mathcal{S}\left(\alpha\right)\mathrm{d}\alpha\,.\label{eq:def-s_n-tilde}\end{equation}
In the following, these coefficients are assumed to be known for
the PAS under consideration. Because $\mathcal{S}\left(\alpha\right)$
must be real, \begin{equation}
\tilde{s}_{n}=\tilde{s}_{-n}^{\ast}\label{eq:FK-hermitesch}\end{equation}
holds. Furthermore we have\begin{equation}
\left|\tilde{s}_{n}\right|\le\tilde{s}_{0}=1\,.\label{eq:Bound-s_n}\end{equation}

A series expansion for $\rho\left(\mathbf{x}\right)$ can be obtained
in a similar way as before for $a\left(\mathbf{x}\right)$: We insert
the Fourier expansion (\ref{eq:FR-PAS}) into Equation (\ref{eq:Def_rho})
and use the integral representation (\ref{eq:Besselsches_Integral-1}).
This yields\begin{equation}
\rho\left(\mathbf{x}\right)=\sum_{n=-\infty}^{\infty}\tilde{s}_{n}\mathrm{e}^{j\beta n}j^{n}\mathrm{J}_{n}\left(2\pi r\right)\,.\label{eq:ACF-Bessel}\end{equation}

The US property (\ref{eq:US-Annahme}) has consequences also for the
correlations between the (random) Fourier coefficients $\hat{a}_{n}$
of $\hat{a}\left(\alpha\right)$. The autocorrelation matrix $\tilde{\mathbf{R}}$
of the discrete random process of the coefficients $\tilde{a}_{n}=2\pi\hat{a}_{n}$
is defined by its elements\begin{equation}
\tilde{R}_{mn}=\mathrm{E}\left\{ \tilde{a}_{m}\tilde{a}_{n}^{\ast}\right\} \,.\label{eq:R_mn_tilde-1}\end{equation}
We insert (\ref{eq:def-s_n-tilde}) into this equation, use the US
assumption (\ref{eq:US-Annahme}), compare with (\ref{eq:def-s_n-tilde})
to obtain the following expression:\begin{equation}
\tilde{R}_{mn}=\tilde{s}_{m-n}\label{eq:R_mn_tilde_FK}\end{equation}
The (infinite) autocorrelation matrix $\tilde{\mathbf{R}}$ can be
written compactly by defining $\tilde{\mathbf{a}}$ as a column vector
of infinite length build from the coefficients $\tilde{a}_{n}$. Then
the autocorrelation matrix can be written as\begin{equation}
\tilde{\mathbf{R}}=\mathrm{E}\left\{ \tilde{\mathbf{a}}\tilde{\mathbf{a}}^{\dagger}\right\} \,.\label{eq:Def_R_tilde}\end{equation}
The dagger $\dagger$ denotes the Hermitian conjucate, i.e. the complex
conjugate transpose of the vector.

\subsection{Time Correlations and Doppler Spectrum}

A single-antenna receiver mounted on a vehicle experiences time variant
fading given by $a\left(\mathbf{x}\left(t\right)\right)$, where
$\mathbf{x}\left(t\right)$ is the antenna position at time $t$.
The time autocorrelation of the fading is given by \begin{equation}
\mathrm{E}\left\{ a\left(\mathbf{x}\left(t\right)\right)a^{\ast}\left(\mathbf{x}\left(t^{\prime}\right)\right)\right\} =\rho\left(\mathbf{x}\left(t\right)-\mathbf{x}\left(t^{\prime}\right)\right)\,.\label{eq:ACF-time-def}\end{equation}
For for $\mathbf{x}\left(t\right)=\mathbf{v}t$, i.e. a vehicle moving
with constant velocity vector $\mathbf{v}$,\begin{equation}
\mathrm{E}\left\{ a\left(\mathbf{x}\left(t\right)\right)a^{\ast}\left(\mathbf{x}\left(t^{\prime}\right)\right)\right\} =\rho_{time}\left(t-t^{\prime}\right)\label{eq:ACF-time-def-1}\end{equation}
holds  with a time autocorrelation function defined by \begin{equation}
\rho_{time}\left(t\right)=\rho\left(\mathbf{v}t\right)\,.\label{eq:rho_time}\end{equation}
Consider a motion with constant velocity $\nu_{max}=\left|\mathbf{v}\right|$
along the x-axis. According to Equation (\ref{eq:k_und_x}), we have
to replace $r\cos\left(\alpha-\beta\right)=\nu_{max}t\cos\left(\alpha\right)$
in Equation (\ref{eq:Def_rho}) which leads to the integral expression
\begin{equation}
\rho_{time}\left(t\right)=\int_{-\pi}^{\pi}\mathrm{e}^{j2\pi\nu_{max}t\cos\left(\alpha\right)}\mathcal{S}\left(\alpha\right)\mathrm{d}\alpha\,.\label{eq:rho_time-1}\end{equation}
We split up this integral according to $\int_{-\pi}^{\pi}\mathrm{d}\alpha=\int_{-\pi}^{0}\mathrm{d}\alpha+\int_{0}^{\pi}\mathrm{d}\alpha$,
substitute $\alpha$ by the Doppler frequency%
\footnote{Note that, because the wavelength is set equal to one, the maximal
Doppler frequency and velocity have the same value $\nu_{max}=\left|\mathbf{v}\right|$. %
} \begin{equation}
\nu=\nu_{max}\cos\left(\alpha\right)\,,\label{eq:Def-nu}\end{equation}
and obtain\begin{equation}
\rho_{time}\left(t\right)=\int_{-\nu_{max}}^{\nu_{max}}\mathrm{e}^{j2\pi\nu t}\mathcal{S}_{Doppler}\left(\nu\right)\mathrm{d}\nu\label{eq:rho_time-2}\end{equation}
with the Doppler spectrum \begin{equation}
\mathcal{S}_{Doppler}\left(\nu\right)=\frac{\mathcal{S}\left(\alpha\left(\nu\right)\right)+\mathcal{S}\left(-\alpha\left(\nu\right)\right)}{\sqrt{\nu_{max}^{2}-\nu^{2}}}\,.\label{eq:S_Doppler-def}\end{equation}
Here we have defined \begin{equation}
\alpha\left(\nu\right)=\arccos\left(\frac{\nu}{\nu_{max}}\right)\,.\label{eq:alpha(nu)}\end{equation}
It is worth to note that Equation (\ref{eq:S_Doppler-def}) reflects
the \emph{mirror symmetry of the Doppler shift}: Two signals impinging
from angles $\alpha$ and $-\alpha$ lead to the same value of $\nu$.
Measurements along a straight line cannot distinguish between signals
from left and right.

Setting $\mathbf{x}=\left(\nu_{max}t,0\right)^{T}$ in Equation (\ref{eq:ACF-Bessel})
yields the series expansion \begin{equation}
\rho_{time}\left(t\right)=\sum_{n=-\infty}^{\infty}\tilde{s}_{n}j^{n}\mathrm{J}_{n}\left(2\pi\nu_{max}t\right)\,.\label{eq:rho_time-Bessel}\end{equation}

The special case of isotropic scattering with PAS \begin{equation}
\mathcal{S}\left(\alpha\right)=\frac{1}{2\pi}\label{eq:isotropes_PAS}\end{equation}
is a popular fading channel model. The corresponding so-called Jakes
\cite{Jakes} Doppler spectrum can be obtained from Equation (\ref{eq:S_Doppler-def})
as\begin{equation}
\mathcal{S}_{Doppler}\left(\nu\right)=\frac{1}{\pi}\frac{1}{\sqrt{\nu_{max}^{2}-\nu^{2}}}\,.\label{eq:Jakes_Doppler}\end{equation}
The corresponding time autocorrelation function is given by\begin{equation}
\rho_{time}\left(t\right)=\mathrm{J}_{0}\left(2\pi\nu_{max}t\right)\,\label{eq:Jakes_ACF}\end{equation}
because in the case of the isotropic PAS the series (\ref{eq:rho_time-Bessel})
reduces to the term with $n=0$. We emphasize that our investigations
are not restricted to the isotropic scattering case.

\section{The Autocorrelation Operator and its Diversity Spectrum}

The goal of this section is to calculate the diversity spectrum of
the KL expansion, i.e. the eigenvalues of the  integral kernel given
by the autocorrelation function of the stochastic process under consideration. 

Consider for example a stochastic process $a\left(t\right)$ with
time variable $t$ restricted to a time interval $\left[t_{1},t_{2}\right]$.
The kernel $R\left(t,t^{\prime}\right)=\mathrm{E}\left\{ a\left(t\right)a^{\ast}\left(t^{\prime}\right)\right\} $
defines a linear operator on the Hilbert space of square integrable
function over the interval $\left[t_{1},t_{2}\right]$. The task is
to calculate its diversity spectrum, which depends on the interval
and on the statistics of $a\left(t\right)$. The KL theory for such
a time interval can be found in Sec. 6-4 and Appendix 2 of \cite{Davenport_Root}.
In the following, we shall need the KL theory for a more general aperture
given by a 1D curve or a 2D region in the plane. We shall define the
Hilbert space of square integrable functions over such a general aperture
and show a method to calculate the corresponding diversity spectrum.

\subsection{Construction of the Hilbert Space}

Consider the following geometrical configurations called \emph{apertures}
and denoted by $\mathcal{A}$: 
\begin{enumerate}
\item Continuous and piecewise smooth curves of finite length in the plane.
Examples are the circle $\mathcal{A}=\left\{ \left(x,y\right)\in\mathbb{R}^{2}:\, x^{2}+y^{2}=1\right\} $
and the interval $\mathcal{A}=\left\{ \left(x,y\right)\in\mathbb{R}^{2}:\,0\le x\le1\mbox{ and }y=0\right\} $. 
\item Arbitrary finite and closed (topologically speaking: compact) two-dimensional
regions in the plane. Examples are the disk $\mathcal{A}=\left\{ \left(x,y\right)\in\mathbb{R}^{2}:\, x^{2}+y^{2}\le1\right\} $
and the square $\mathcal{A}=\left\{ \left(x,y\right)\in\mathbb{R}^{2}:\, x,\, y\le1/2\right\} $.
\end{enumerate}
Both cases can be be related to an idealised antenna array with infinitely
dense antenna packing on the curve or region, respectively. We shall
analyse the multipath richness of the stochastic process $a\left(\mathbf{x}\right)$
resticted to that aperture. This has to be done by solving the eigenvalue
problem of the autocorrelation integral kernel which is related to
the KL expansion on that aperture. The first task is to construct
the Hilbert space $\mathcal{H}=\mathcal{L}^{2}\left(\mathcal{A},\mathrm{d}\mu\right)$
of square-integrable functions $\phi,\,\psi,\,...$ that live on that
aperture $\mathcal{A}$:\begin{equation}
\int_{\mathcal{A}}\,\left|\phi\right|^{2}\mathrm{d}\mu<\infty\label{eq:quadratintegrabel}\end{equation}
Here $\mu$ is the (normalised) Lebesgue measure on $\mathcal{A}$.
The scalar product of two vectors $\phi,\,\psi\in\mathcal{H}$ is
given by \begin{eqnarray}
\left\langle \phi,\psi\right\rangle  & = & \int_{\mathcal{A}}\,\phi^{\ast}\psi\mathrm{d}\mu\label{eq:Definition_Skalarprodukt-allgemein}\end{eqnarray}
and the norm of $\phi\in\mathcal{H}$ is defined by \begin{equation}
\left\Vert \phi\right\Vert =\sqrt{\left\langle \phi,\phi\right\rangle }\,.\label{eq:Definition_HR-Norm}\end{equation}

To obtain the Karhunen-Loève (KL) expansion, we need to define the
autocorrelation operator $\mathbf{R}$ on that Hilbert space. For
convenience, we have normalised the measure to \begin{equation}
\mu\left(\mathcal{A}\right)\triangleq\int_{\mathcal{A}}\mathrm{d}\mu=1\,.\label{eq:Normalised_Measure}\end{equation}
As we shall see, this has the consequence that the trace $\mathrm{tr}\left(\mathbf{R}\right)$
(i.e. the sum of eigenvalues) of the operator $\mathbf{R}$ is normalised
to one. 

Some facts about Hilbert space operators and definitions of their
different norms are summarised in Appendix A.

\subsubsection{The Hilbert Space for a One-Dimensional Curve}

Consider a finite-length, continuous, and piecewise smooth curve $\mathcal{A}$
in the plane. For a given parametrisation $\tau\mapsto\mathbf{x}\left(\tau\right)$
the curve is defined by \[
\mathcal{A}=\left\{ \mathbf{x}\left(\tau\right)\in\mathbb{R}^{2}:\,\tau_{1}\le\tau\le\tau_{2}\right\} \,,\]
and $\mathcal{A}$ is just the image of an interval $\left[\tau_{1},\tau_{2}\right]$.
Thus, the Hilbert space of square integrable functions over $\mathcal{A}$
is isomorphic to the Hilbert space of square integrable functions
over that interval. In the following, this Hilbert space will be constructed
explicitely. The line integral of a scalar function $\Upsilon$ over
a curve $\mathcal{A}$ is defined as\begin{equation}
\int_{\mathcal{A}}\Upsilon\mathrm{d}\ell=\int_{\tau_{1}}^{\tau_{2}}\Upsilon\left(\mathbf{x}\left(\tau\right)\right)\left|\dot{\mathbf{x}}\left(\tau\right)\right|\mathrm{d}\tau\,.\label{eq:Linienintegral}\end{equation}
In that equation, $\mathbf{x}\left(\tau\right)$ is a piecewise differentiable
paramatrisation of the curve $\mathcal{A}$, and $\dot{\mathbf{x}}\left(\tau\right)$
stands for the derivative with respect to the parameter $\tau\in\left[\tau_{1},\tau_{2}\right]$.
The integral (\ref{eq:Linienintegral}) is independent of the choice
of the parametrization. The length of the curve can be expressed by
the curve integral over the scalar function $\Upsilon=1$ as \begin{equation}
\left|\mathcal{A}\right|=\int_{\mathcal{A}}\mathrm{d}\ell\,.\label{eq:Laenge_der_Kurve}\end{equation}
A special and frequently preferred parameter is the (running) length
$\lambda$ of the curve. In that case $\left|\dot{\mathbf{x}}\left(\lambda\right)\right|=1$
holds, and Equation (\ref{eq:Linienintegral}) becomes \begin{equation}
\int_{\mathcal{A}}\Upsilon\mathrm{d}\ell=\int_{0}^{\left|\mathcal{A}\right|}\Upsilon\left(\mathbf{x}\left(\lambda\right)\right)\mathrm{d}\lambda\label{eq:Linienintegral-Laenge}\end{equation}
For this paper, we will chose $\tau$ to be the \emph{normalised length}
\begin{equation}
\tau=\frac{\lambda}{\left|\mathcal{A}\right|}\,.\label{eq:Normierte_Laenge}\end{equation}
For this so-called \emph{priviledged} parameter $\tau$ we obtain
(by using a simple substitution)\begin{equation}
\frac{1}{\left|\mathcal{A}\right|}\int_{\mathcal{A}}\Upsilon\mathrm{d}\ell=\int_{0}^{1}\Upsilon\left(\tau\right)\mathrm{d}\tau\,.\label{eq:Linienintegral-01}\end{equation}
In that equation, we have written $\Upsilon\left(\tau\right)\triangleq\Upsilon\left(\mathbf{x}\left(\tau\right)\right)\triangleq\Upsilon\left(\mathbf{x}\left(\lambda\left(\tau\right)\right)\right)$
to simplify the notation. We shall use this loose notation whenever
it does not give rise to any confusion. Equation (\ref{eq:Linienintegral-01})
means that we can express every average over the curve by an average
over the unit interval. 

Bei chosing the priveledged parameter, the Hilbert space becomes isomorhic
to the Hilbert space over the unit interval: \begin{equation}
\mathcal{H}=\mathcal{L}^{2}\left(\mathcal{A},\mu\right)\cong\mathcal{L}^{2}\left(\left[0,1\right],\mathrm{d}\tau\right)\,\label{eq:Def_Hilbertraum-Kurve}\end{equation}
Here $\cong$ denotes isomorphism, the measure $\mu$ is given by
\begin{equation}
\mathrm{d}\mu=\frac{\mathrm{d}\ell}{\left|\mathcal{A}\right|}\,,\label{eq:d_mu_curve}\end{equation}
and the scalar product (\ref{eq:Definition_Skalarprodukt-allgemein})
can be written in explicit form as\begin{eqnarray}
\left\langle \phi,\psi\right\rangle  & = & \int_{0}^{1}\phi^{\ast}\left(\tau\right)\psi\left(\tau\right)\mathrm{d}\tau\,.\label{eq:Definition_Skalarprodukt-KUrve}\end{eqnarray}

\paragraph{Example 1: The Hilbert space of functions on a circle }

We consider a circle of radius $r$ defined by\begin{equation}
\mathcal{A}=\left\{ \mathbf{x}\in\mathbb{R}^{2}:\,\left|\mathbf{x}\right|=r\right\} \,.\label{eq:Def_Circle}\end{equation}
The priviledged parametrisation is given by\begin{equation}
\mathbf{x}\left(\tau\right)=r\left(\cos\left(2\pi\tau\right),\,\sin\left(2\pi\tau\right)\right)^{T},\,\label{eq:Parameter_Circle}\end{equation}
and $\left|\mathcal{A}\right|=2\pi r$. The scalar product \begin{equation}
\left\langle \phi,\psi\right\rangle =\frac{1}{2\pi r}\int_{\mathcal{A}}\,\phi^{\ast}\psi\mathrm{d}\ell=\int_{0}^{1}\phi^{\ast}\left(\tau\right)\psi\left(\tau\right)\mathrm{d}\tau\,.\label{eq:Skalarprodukt_Circle}\end{equation}
is just the standard scalar product on the unit circle. We note that
the natural orthonormal%
\footnote{The basis $\left\{ u_{n}\right\} $ is called orthonormal if $\left\langle u_{m},u_{n}\right\rangle =\delta_{mn}$
holds.%
} basis on the unit circle is the Fourier basis $\left\{ u_{n}\right\} _{n=-\infty}^{\infty}$
with \begin{equation}
u_{n}\left(\tau\right)=\mathrm{e}^{j2\pi\tau n}\,.\label{eq:Fourierbasis}\end{equation}
Assuming that each realisation of the fading is an element of the
Hilbert space, we can expand it with respect to this basis\begin{equation}
a\left(\tau\right)=\sum_{n=-\infty}^{\infty}a_{n}u_{n}\left(\tau\right)\label{eq:Fourierentwicklung_Fading}\end{equation}
with coefficients given by\begin{equation}
a_{n}=\left\langle u_{n},a\right\rangle \,.\label{eq:Fourierkoeff_a_n}\end{equation}
This is the familiar formula for the Fourier coefficients: \begin{eqnarray}
a_{n} & = & \int_{0}^{1}u_{n}^{\ast}\left(\tau\right)a\left(\tau\right)\mathrm{d}\tau\label{eq:Fourierkoeff-a_n-1}\\
 & = & \int_{0}^{1}\mathrm{e}^{-j2\pi\tau n}a\left(\tau\right)\mathrm{d}\tau\end{eqnarray}

\subsubsection{The Hilbert Space for a Two-Dimensional Region}

We now consider consider a closed two-dimensional region $\mathcal{A}\subset\mathbb{R}^{2}$
of finite area $\left|\mathcal{A}\right|$. The functions $\phi\left(\mathbf{x}\right)$
that satisfy the condition \[
\int_{\mathcal{A}}\left|\phi\left(\mathbf{x}\right)\right|^{2}\mathrm{d}^{2}\mathbf{x}=<\infty\,\]
build the Hilbert space \begin{equation}
\mathcal{H}=\mathcal{L}^{2}\left(\mathcal{A},\mu\right)\,,\label{eq:Def_Hilbertraum-Area}\end{equation}

with the measure $\mu$ given by \begin{equation}
\mathrm{d}\mu=\frac{\mathrm{d}^{2}\mathbf{x}}{\left|\mathcal{A}\right|}\,.\label{eq:d_mu_area}\end{equation}
The scalar product of two vector $\phi\,,\psi\in\mathcal{H}$ is defined
by \begin{eqnarray}
\left\langle \phi,\psi\right\rangle  & = & \int_{\mathcal{A}}\,\phi^{\ast}\left(\mathbf{x}\right)\psi\left(\mathbf{x}\right)\mathrm{d}\mu\left(\mathbf{x}\right)\label{eq:Definition_Skalarprodukt-Region}\\
 & = & \frac{1}{\left|\mathcal{A}\right|}\int_{\mathcal{A}}\,\phi^{\ast}\left(\mathbf{x}\right)\psi\left(\mathbf{x}\right)\mathrm{d^{2}}\mathbf{x}\,.\end{eqnarray}

\paragraph{Example 2: The Hilbert space of functions on a disk \cite{Kennedy2007}}

We consider a close disk of radius $r_{1}$, i.e.\begin{equation}
\mathcal{A}=\left\{ \mathbf{x}\in\mathbb{R}^{2}:\,\left|\mathbf{x}\right|\le r_{1}\right\} \,.\label{eq:Def_disk}\end{equation}
It has the area $\left|\mathcal{A}\right|=\pi r_{1}^{2}.$ The scalar
product is defined by\begin{equation}
\left\langle \phi,\psi\right\rangle =\frac{1}{\left|\mathcal{A}\right|}\int_{\mathcal{C}}\,\phi\left(\mathbf{x}\right)^{\ast}\psi\left(\mathbf{x}\right)\mathrm{d}^{2}\mathbf{x}\,.\label{eq:Skalarprodukt_disk}\end{equation}
Introducing polar coordinates $r$ and $\beta$ according to Equation
(\ref{eq:k_und_x}) yields\[
\frac{\mathrm{d}^{2}\mathbf{x}}{\left|\mathcal{A}\right|}=\frac{\mathrm{d}\beta\, r\mathrm{d}r}{\pi r_{1}^{2}}\,.\]
A we shall see later, an orthonormal basis for this Hilbert space
on a disk is given by\[
u_{n}\left(\mathbf{x}\right)=\frac{\mathrm{J}_{n}\left(2\pi\left|\mathbf{x}\right|\right)}{\sqrt{\int_{0}^{r_{1}}\frac{2\pi r\mathrm{d}r}{\left|\mathcal{A}\right|}\,\mathrm{J}_{n}^{2}\left(2\pi r\right)}}\mathrm{e}^{j\beta\left(\mathbf{x}\right)n}\,.\]

\subsection{The Autocorrelation Operator }

We now define the autocorrelation operator $\mathbf{R}$ on the one-
or two-dimensional aperture $\mathcal{A}$.

\paragraph{One-dimensional aperture: }

We define the operator $\mathbf{R}$ acting on $\mathcal{H}\cong\mathcal{L}^{2}\left(\left[0,1\right]\right)$
by its integral kernel\begin{equation}
R\left(\tau,\tau^{\prime}\right)=\rho\left(\mathbf{x}\left(\tau\right)-\mathbf{x}\left(\tau^{\prime}\right)\right)\,\label{eq:Kernel_tau}\end{equation}
with $\rho\left(\mathbf{x}\right)$ according to Equations (\ref{eq:AKF-1})
and (\ref{eq:Def_rho}) and $\mathbf{x}\left(\tau\right)$ being the
priviledged parametrisation of the curve. Using loose notation $a\left(\tau\right)\triangleq a\left(\mathbf{x}\left(\tau\right)\right)$
in Equation (\ref{eq:AKF-1}) we may write \begin{equation}
R\left(\tau,\tau^{\prime}\right)=\mathrm{E}\left\{ a\left(\tau\right)a^{\ast}\left(\tau^{\prime}\right)\right\} \,.\label{eq:Kernel_tau-1}\end{equation}
The action of the autocorrelation operator $\mathbf{R}$ on a vector
$\psi\in\mathcal{H}$ is given by\begin{equation}
\left(\mathbf{R}\psi\right)\left(\tau\right)=\int_{0}^{1}R\left(\tau,\tau^{\prime}\right)\psi\left(\tau^{\prime}\right)\mathrm{d}\tau^{\prime}\,.\label{eq:Wirkung_R-Kurve}\end{equation}
We note that the kernel is \emph{not} a convolution kernel (i.e. a
kernel that depends only on the difference variable $\tau-\tau^{\prime}$)
except for the special case where $\mathcal{A}$ is a straight line.

\paragraph{Two-dimensional aperture: }

We define the operator $\mathbf{R}$ acting on $\mathcal{H}=\mathcal{L}^{2}\left(\mathcal{A},\mu\right)$
by its integral kernel \begin{equation}
R\left(\mathbf{x},\mathbf{x}^{\prime}\right)=\rho\left(\mathbf{x}-\mathbf{x}^{\prime}\right)\,,\label{eq:Kernel_x}\end{equation}
defined for $\mathbf{x},\,\mathbf{x}^{\prime}\in\mathcal{A}$. Again,
$\rho\left(\mathbf{x}\right)$ is given by Equations (\ref{eq:AKF-1})
and (\ref{eq:Def_rho}). The action of the autocorrelation operator
$\mathbf{R}$ on a vector $\psi\in\mathcal{H}$ is given by\begin{equation}
\left(\mathbf{R}\psi\right)\left(\tau\right)=\frac{1}{\left|\mathcal{A}\right|}\int_{\mathcal{A}}R\left(\mathbf{x},\mathbf{x}^{\prime}\right)\psi\left(\mathbf{x}^{\prime}\right)\mathrm{d}^{2}\mathbf{x}\,.\label{eq:Wirkung_R-Flaeche}\end{equation}

In the following treatment, we will no further distinguish between
both cases. One can show by direct calculation from Equation (\ref{eq:Wirkung_R-Kurve})
or (\ref{eq:Wirkung_R-Flaeche}) that\begin{equation}
\left\langle \phi,\mathbf{R}\psi\right\rangle =\mathrm{E}\left\{ \left\langle \phi,a\right\rangle \left\langle a,\psi\right\rangle \right\} \,.\label{eq:Def_R-1}\end{equation}
holds. The values of these so-called \emph{matrix elements} $\left\langle \phi,\mathbf{R}\psi\right\rangle $
uniquely define the linear operator $\mathbf{R}$ \cite{Reed-Simon-1}.
Thus, Equation (\ref{eq:Def_R-1}) can be interpreted as an alternative
definition of the operator that is valid for both classes of apertures.

We list some properties of the autocorrelation operator $\mathbf{R}$:

\paragraph{Continuity, integrability, and compactness:}

From Lebesgue's dominated convergence theorem and the absolute integrability
of $\mathcal{S}\left(\alpha\right)$ it follows that $\rho\left(\mathbf{x}\right)$
given by integral expression (\ref{eq:Def_rho}) is a continuous function
on $\mathcal{A}$. Thus, the kernel $R\left(\tau,\tau^{\prime}\right)$
(or $R\left(\mathbf{x},\mathbf{x}^{\prime}\right)$, respectively)
is a continuous function in both variables. Because any continuous
function on a closed, finite set is integrable (and square-integrable,
too) the kernel is in $\mathcal{L}^{2}\left(\mathcal{A},\mathrm{d}\mu\right)\times\mathcal{L}^{2}\left(\mathcal{A},\mathrm{d}\mu\right)$.
This property ensures that $\mathbf{R}$ is a Hilbert-Schmidt operator,
which also guaranties that it is compact operator (see Theorems VI
22 and VI 23 in \cite{Reed-Simon-1}). Because $\left|\rho\left(\mathbf{x}\right)\right|\le1$
it can further be shown that the operator norm of $\mathbf{R}$ is
bounded by $\left\Vert \mathbf{R}\right\Vert \le1$.

\paragraph{Symmetry and positive semi-definiteness: }

Because the operator $\mathbf{R}$ is bounded, it is defined on the
whole Hilbert space. Because of Equation (\ref{eq:Kernel_tau-1}),
the kernel is symmetric\begin{equation}
R\left(\tau,\tau^{\prime}\right)=R^{\ast}\left(\tau^{\prime},\tau\right)\,,\label{eq:kernel-symmetric}\end{equation}
and the corresponding operator $\mathbf{R}$ is self-adjoint%
\footnote{See Appendix A for the definition of self-adjointness.%
}:\begin{equation}
\mathbf{R}=\mathbf{R}^{\dagger}\label{eq:selfadjoint}\end{equation}
The same hold with $R\left(\tau,\tau^{\prime}\right)$ replaced by
$R\left(\mathbf{x},\mathbf{x}^{\prime}\right)$. From Equation (\ref{eq:Def_R-1}),
we conclude that $\mathbf{R}$ is positive semi-definite, i.e. \[
\left\langle \psi,\mathbf{R}\psi\right\rangle \ge0\]
holds for every $\psi\in\mathcal{H}$. We also say that the corresponding
kernel $R\left(\cdot,\cdot\right)$ is positiv semidefinite.

\paragraph{Eigenvalues and eigenvectors of $\mathbf{R}$:}

From the properties of $\mathbf{R}$ stated above, the following properties
of its eigenvalues can be infered (\cite{Reed-Simon-1}, Chap. VI):
Since $\mathbf{R}$ is self-adjoint, there is a complete set of eigenvectors
of $\mathbf{R}$, $\left\{ \varphi_{i}\right\} _{i=1}^{\infty}$,
with \begin{equation}
\left\langle \varphi_{i},\varphi_{k}\right\rangle =\delta_{ik}\label{eq:orthonormal}\end{equation}
that build a basis of $\mathcal{H}$. Because\textbf{ $\mathbf{R}$}
is compact, the corresponding eigenvalues $\mathrm{eig}\left(\mathbf{R}\right)\triangleq\left\{ \lambda_{i}\right\} _{i=1}^{\infty}$
are of finite multiplicity and have the property \[
\lim_{i\to\infty}\lambda_{i}=0\,.\]
Because the operator is positive semidefinite, we know that $\lambda_{i}\ge0$
for all $i$.

\paragraph{Mercer's Theorem: }

Because the kernel $R\left(\tau,\tau^{\prime}\right)$ is continuous,
symmetric, and positive semidefinite, according to Mercer's Theorem
(\cite{Courant_Hilbert-1}, §5.4; \cite{Riesz-Nagy}, § 98) the following
series expansion for the kernel holds: \begin{equation}
R\left(\tau,\tau^{\prime}\right)=\sum_{i=1}^{\infty}\lambda_{i}\varphi_{i}\left(\tau\right)\varphi_{i}^{\ast}\left(\tau^{\prime}\right)\,\label{eq:Mercer-1}\end{equation}
The theorem states that the infinite series converges uniformly in
$\tau$ and $\tau^{\prime}$. For the kernel $R\left(\mathbf{x},\mathbf{x}^{\prime}\right)$,
the analogous expansion hold with $\tau,\,\tau^{\prime}$ replaced
by $\mathbf{x},\,\mathbf{x}^{\prime}$.

\paragraph{The trace of the operator:}

From Mercer's theorem we conclude\begin{eqnarray}
\int_{0}^{1}R\left(\tau,\tau\right)\mathrm{d}\tau & = & \sum_{i=1}^{\infty}\lambda_{i}\int_{0}^{1}\varphi_{i}\left(\tau\right)\varphi_{i}^{\ast}\left(\tau\right)\mathrm{d}\tau\nonumber \\
\int_{0}^{1}R\left(\tau,\tau\right)\mathrm{d}\tau & = & \sum_{i=1}^{\infty}\lambda_{i}\label{eq:Spur_als_Integral}\end{eqnarray}
Due to our normalisation $\rho\left(0\right)=1$ we have $R\left(\tau,\tau\right)=\rho\left(0\right)=1$
and, thus,\begin{equation}
\mathrm{tr}\left(\mathbf{R}\right)=\sum_{i=1}^{\infty}\lambda_{i}=1\,.\label{eq:Spur.eq.1}\end{equation}
The same holds for $\tau$ replaced by $\mathbf{x}$.

\paragraph{Karhunen-Loève Theorem: }

Let $\left\{ \varphi_{n}\right\} _{n=1}^{\infty}$ be the basis of
eigenvectors of $\mathbf{R}$ and $\mathrm{eig}\left(\mathbf{R}\right)=\left\{ \lambda_{n}\right\} _{n=1}^{\infty}$
be the corresponding eigenvalues defined by \begin{equation}
\mathbf{R}\varphi_{n}=\lambda_{n}\varphi_{n}\,.\label{eq:EW-Gl}\end{equation}
Define the (random) coefficients\begin{equation}
c_{n}=\left\langle \varphi_{n},a\right\rangle =\int_{0}^{1}\varphi_{n}^{\ast}\left(\tau\right)a\left(\tau\right)\mathrm{d}\tau\,.\label{eq:c_i}\end{equation}
One can readily concluse from Equation (\ref{eq:EW-Gl}) that these
coefficients must be uncorrelated:\begin{equation}
\mathrm{E}\left\{ c_{m}c_{n}^{\ast}\right\} =\delta_{mn}\lambda_{n}\,.\label{eq:c_n_uncorr}\end{equation}
Conversely, one can see that the basis of eigenvectors $\varphi_{n}$
of $\mathbf{R}$ is the only one such that the coefficients (\ref{eq:c_i})
fulfil the condition (\ref{eq:c_n_uncorr}). The KL theorem states
that the expansion\begin{equation}
a\left(\tau\right)=\sum_{n=1}^{\infty}c_{n}\varphi_{n}\left(\tau\right)\label{eq:KLE-2}\end{equation}
converges as a limit in mean (l.i.m.) and that the convergence is
uniform in $\tau$. The same holds for the variable $\tau$ replaced
by $\mathbf{x}$. It is worth to note that the proof of convergence
(see \cite{Van_Trees,Davenport_Root}) is based on Mercer's theorem
which requires the continuity of the kernel.

\paragraph{Diversity Measure:}

We refer to Appendix A for the definition of the norms $\bigl\Vert\mathbf{A}\bigr\Vert$,
$\bigl\Vert\mathbf{A}\bigr\Vert_{1}$, and $\bigl\Vert\mathbf{A}\bigr\Vert_{2}$
of an operator $\mathbf{A}$. By using these norms, the diversity
measure given by Equation (\ref{eq:Diversity-Measure}) can be re-defined
as \begin{equation}
\omega\left(\mathbf{R}\right)=\frac{\bigl\Vert\mathbf{R}\bigr\Vert_{1}^{2}}{\bigl\Vert\mathbf{R}\bigr\Vert_{2}^{2}}\,,\label{eq:Diversity_measure-3}\end{equation}
i. e. as the ratio between the trace norm \begin{equation}
\bigl\Vert\mathbf{R}\bigr\Vert_{1}=\sum_{n=1}^{\infty}\lambda_{n}\label{eq:}\end{equation}
and the Hilbert-Schmidt norm\begin{equation}
\bigl\Vert\mathbf{R}\bigr\Vert_{2}=\sqrt{\sum_{n=1}^{\infty}\lambda_{n}^{2}}\label{eq:-1}\end{equation}
of the operator $\mathbf{R}$. Because of hierarchy of norm given
by Inequality (\ref{eq:normhierarchie}), $\omega\left(\mathbf{R}\right)\ge1$
always holds.

Equation (\ref{eq:Diversity_measure-3}) is a suitable definition
of the diversity measure for an integral kernel as studied in this
paper. The equivalent for a finite-dimensional $L\times L$ correlation
matrix $\mathbf{R}$ it has been defined in \cite{Ivrlac2003,Muharemovic2008}.
In the next section we shall show numerically that for $L$ antennas
distributed uniformly over a given aperture, the calculated diversity
measure approaches that of a dense array in the limit $L\to\infty$. 

The value of the quantity $\omega$ can be understood by considering
two extreme cases for a finite number of antennas \cite{Ivrlac2003,Muharemovic2008}:
Firstly, for $L$ uncorrelated antennas, $\mathbf{R}$ is equal to
the identity matrix $\mathbf{I}$ with $\bigl\Vert\mathbf{I}\bigr\Vert_{1}^{2}=L^{2}$
and $\bigl\Vert\mathbf{I}\bigr\Vert_{2}^{2}=L$, which yields $\omega=L$.
Secondly, for $L$ antennas with 100\% correlation, $R_{ik}=1$ holds
for all elements $R_{ik}$ of $\mathbf{R}$. This means that $\bigl\Vert\mathbf{R}\bigr\Vert_{1}^{2}=L^{2}$
and $\bigl\Vert\mathbf{R}\bigr\Vert_{2}^{2}=L^{2}$, which yields
$\omega=1$. Thus, the diversity measure $\omega\left(\mathbf{R}\right)$
can be interpreted as the number of equivalent uncorrelated antennas.
Note, however, that $\omega\left(\mathbf{R}\right)$ is typically
not an integer number.

For the continuous case, due to the trace normalisation (\ref{eq:Spur.eq.1}),
we have simply\begin{equation}
\omega\left(\mathbf{R}\right)=\frac{1}{\bigl\Vert\mathbf{R}\bigr\Vert_{2}^{2}}\ge1\,.\label{eq:Diversity_measure-4}\end{equation}

\subsection{Discrete Representation of $\mathbf{R}$}

To calculate the diversity spectrum $\mathrm{eig}\left(\mathbf{R}\right)=\left\{ \lambda_{n}\right\} _{n=1}^{\infty}$,
one must solve the eigenvalue problem given by Equation (\ref{eq:EW-Gl}).
Expressing $\mathbf{R}$ by its integral kernel yields integral equation\begin{equation}
\int_{0}^{1}R\left(\tau,\tau^{\prime}\right)\varphi_{n}\left(\tau^{\prime}\right)\mathrm{d}\tau^{\prime}=\lambda_{n}\varphi_{n}\left(\tau\right)\,\label{eq:EW-Gl-Integral}\end{equation}
(or the corresponding expression in the variable $\mathbf{x}$). This
may be solved by numerical quadrature methods. However, this approach
is numerically cumbersome, and the accuracy is difficult to control.
An alternative approach introduced by \cite{Kennedy2007} is to switch
from the continuous representation given by the integral kernel $R\left(\cdot,\cdot\right)$
to a discrete representation corresponding to an orthonormal basis
$\left\{ u_{n}\right\} $ of $\mathcal{H}$. Then the equivalent
matrix%
\footnote{The slightly lose notation to use the same symbol for the operator
and its equivalent matrix representation does not cause problems,
and it is quite usual e.g. in quantum mechanics.%
} $\mathbf{R}$ with matrix elements given by\begin{equation}
R_{mn}=\left\langle u_{m},\mathbf{R}u_{n}\right\rangle \,.\label{eq:R_mn}\end{equation}
in this new representation has the same eigenvalues. These initially
unknown matrix elements have to be obtained from the known matrix
elements $\tilde{R}_{mn}$ defined by Equation (\ref{eq:R_mn_tilde_FK}).
From Equation (\ref{eq:Def_R-1}) we find that\begin{equation}
R_{mn}=\mathrm{E}\left\{ a_{m}a_{n}^{\ast}\right\} \,\label{eq:R_mn-1-2}\end{equation}
holds, where\begin{equation}
a_{n}=\left\langle u_{n},a\right\rangle \label{eq:a_n}\end{equation}
is the coefficient of the expansion\begin{equation}
a=\sum_{n}a_{n}u_{n}\label{eq:Entwicklung_a_nach_u_n}\end{equation}
of the fading process with respect to this basis. We define the column
vector $\mathbf{a}$ of infinite length that is build from the coefficients
$a_{n}$. The autocorrelation matrix that has $R_{mn}$ as its elements
is then given by \begin{equation}
\mathbf{R}=\mathrm{E}\left\{ \mathbf{a}\mathbf{a}^{\dagger}\right\} \,.\label{eq:Def_R_Matrix}\end{equation}
The eigenvalue problem can this way been reduced to a discrete (but
infinite-dimensional) one. Before soving it, the first task is to
define the appropriate basis $\left\{ u_{n}\right\} $ and then relate
the matrix $\mathbf{R}$ to the matrix $\tilde{\mathbf{R}}$ with
elements given by Equation (\ref{eq:R_mn_tilde_FK}). 

To do this, we expand the fading process $a$ according to Equation
(\ref{eq:a(x)_Bessel-1}) by\begin{equation}
a=\sum_{n=-\infty}^{\infty}\tilde{a}_{n}v_{n}\,,\label{eq:Entwicklung_v_n-1}\end{equation}
where $v_{n}\left(\mathbf{x}\right)$ is defined by \begin{equation}
v_{n}\left(\mathbf{x}\right)=\mathrm{e}^{j\beta\left(\mathbf{x}\right)n}j^{n}\mathrm{J}_{n}\left(2\pi\left|\mathbf{x}\right|\right)\,.\label{eq:Def_v_n(x)}\end{equation}
The domain of the functions $a\left(\mathbf{x}\right)$ and $v_{n}\left(\mathbf{x}\right)$
is the array $\mathcal{A}$, which is either a two-dimensional region
or a one-dimensional curve. For a curve with priviledged parameter
$\tau$ we may write \begin{equation}
v_{n}\left(\tau\right)=\mathrm{e}^{j\beta\left(\tau\right)n}j^{n}\mathrm{J}_{n}\left(2\pi r\left(\tau\right)\right)\label{eq:Def_v_n}\end{equation}
where $\beta\left(\tau\right)=\beta\left(\mathbf{x}\left(\tau\right)\right)$
and $r\left(\tau\right)=\left|\mathbf{x}\left(\tau\right)\right|$
are the polar coordinates of the parametrisation. In general, the
set $\left\{ v_{n}\right\} $ is not orthogonal. For the special case
where $\mathcal{A}$ is a interval on the x-axis, the functions $v_{n}\left(x\right)=j^{n}\mathrm{J}_{n}\left(2\pi x\right)$
do not even form a basis. This is because of the Bessel function property
$\mathrm{J}_{n}\left(2\pi x\right)=\left(-1\right)^{n}\mathrm{J}_{-n}\left(2\pi x\right)$,
they are not linearly independent. But in any case, Equation (\ref{eq:Entwicklung_v_n-1})
holds for every realisation of the fading process restricted on the
array. Each $v_{n}$ can be expanded into the orthonormal basis $\left\{ u_{n}\right\} $
by\begin{equation}
v_{n}=\sum_{m}T_{mn}u_{m}\label{eq:v_n_entw_nach_u_m}\end{equation}
with coefficients%
\footnote{In case that $\left\{ v_{n}\right\} $ is a basis, Equation (\ref{eq:T_mn})
defines an basis transform operator $\mathbf{T}$ by \[
\mathbf{T}:\, u_{n}\mapsto v_{n}=\mathbf{T}u_{n}\,.\]
It has matrix elements \[
T_{mn}=\left\langle u_{m},\mathbf{T}u_{m}\right\rangle =\left\langle u_{m},v_{n}\right\rangle \,.\]
}\begin{equation}
T_{mn}=\left\langle u_{m},v_{n}\right\rangle \,.\label{eq:T_mn}\end{equation}
Expanding\begin{eqnarray*}
a_{m} & = & \left\langle u_{m},a\right\rangle \\
 & = & \left\langle u_{m},\sum_{n}\tilde{a}_{n}v_{n}\right\rangle \end{eqnarray*}
yields\begin{equation}
a_{m}=\sum_{n}T_{mn}\tilde{a}_{n}\,.\label{eq:Transf_Koeff-1}\end{equation}
Let $\mathbf{T}$ be the matrix with elements $T_{mn}$. Then the
above equation can be written compactly as\begin{equation}
\mathbf{a}=\mathbf{T}\tilde{\mathbf{a}}\,.\label{eq:Transf_Koeff-2}\end{equation}
We insert this into the autocorrelation matrix\begin{eqnarray*}
\mathbf{R} & = & \mathrm{E}\left\{ \mathbf{a}\mathbf{a}^{\dagger}\right\} \\
 & = & \mathrm{E}\left\{ \mathbf{T}\tilde{\mathbf{a}}\left(\mathbf{T}\tilde{\mathbf{a}}\right)^{\dagger}\right\} \\
 & = & \mathbf{T}\mathrm{E}\left\{ \tilde{\mathbf{a}}\tilde{\mathbf{a}}^{\dagger}\right\} \mathbf{T}^{\dagger}\end{eqnarray*}
and obtain\begin{equation}
\mathbf{R}=\mathbf{T}\tilde{\mathbf{R}}\mathbf{T}^{\dagger}\,.\label{eq:Transf_Matrizen}\end{equation}
Making use of the property $\mathrm{eig}\left(\mathbf{AB}\right)=\mathrm{eig}\left(\mathbf{BA}\right)$
(see \cite{Horn-Johnson}, p. 53), we obtain \begin{eqnarray*}
\mathrm{eig}\left(\mathbf{R}\right) & = & \mathrm{eig}\left(\mathbf{T}\tilde{\mathbf{R}}\mathbf{T}^{\dagger}\right)\\
 & = & \mathrm{eig}\left(\mathbf{T}^{\dagger}\mathbf{T}\tilde{\mathbf{R}}\right)\end{eqnarray*}
i.e. \begin{equation}
\mathrm{eig}\left(\mathbf{R}\right)=\mathrm{eig}\left(\mathbf{G}\tilde{\mathbf{R}}\right)\label{eq:eig(R)-Formel}\end{equation}
with the Gram matrix $\mathbf{G}$ that is defined by \begin{equation}
\mathbf{G}=\mathbf{T}^{\dagger}\mathbf{T}\,.\label{eq:Gram-Operator}\end{equation}
It has the matrix elements\begin{equation}
G_{mn}=\left\langle v_{m},v_{n}\right\rangle \,.\label{eq:Gram-Matrix}\end{equation}
Since $\mathbf{G}$ and $\tilde{\mathbf{R}}$ are known matrices,
the eigenvalue problem for $\mathbf{R}$ can be solved by utilising
Equation (\ref{eq:eig(R)-Formel}). Before we show how this can be
done with a controlled approximation, we discuss the two examples
from above for which the transform turns out to be quite simple.

\paragraph{Example 1 (ctd.): }

Consider the circle (\ref{eq:Def_Circle}) of radius $r$ with parametrisation
given by (\ref{eq:Parameter_Circle}). According to Equation (\ref{eq:a(x)_Bessel-1}),
the fading process $a\left(\tau\right)$ can be written as\begin{equation}
a\left(\tau\right)=\sum_{n=-\infty}^{\infty}\tilde{a}_{n}j^{n}\mathrm{J}_{n}\left(2\pi r\right)\mathrm{e}^{j2\pi\tau n}\,.\label{eq:a(tau)}\end{equation}
This is an expansion\begin{equation}
a\left(\tau\right)=\sum_{n=-\infty}^{\infty}\tilde{a}_{n}v_{n}\left(\tau\right)\label{eq:Entwicklung_v_n}\end{equation}
with respect to the basis\begin{equation}
v_{n}\left(\tau\right)=j^{n}\mathrm{J}_{n}\left(2\pi r\right)\mathrm{e}^{j2\pi\tau n}\,.\label{eq:v_n(tau)}\end{equation}
This basis is already orthogonal. Is can be normalised by dividing
it by a normalisation factor $\xi_{n}$ that fullfills the requirement\[
\left|\xi_{n}\right|=\left\Vert v_{n}\right\Vert \,.\]
We define\begin{equation}
\xi_{n}=j^{n}\mathrm{J}_{n}\left(2\pi r\right)\,.\label{eq:xi_n-circle}\end{equation}
We have included the phase factor $j^{n}$ because the basis $\left\{ u_{n}\right\} _{n=-\infty}^{\infty}$
with\begin{equation}
v_{n}\left(\tau\right)=\xi_{n}u_{n}\left(\tau\right)\label{eq:v_n_vs_u_n}\end{equation}
is now just the standard orthonormal Fourier basis $u_{n}\left(\tau\right)=\mathrm{e}^{j2\pi\tau n}$. 

Inserting into the expansion (\ref{eq:Entwicklung_v_n}) and comparing
with Equation (\ref{eq:Fourierentwicklung_Fading}), we find that
the corresponding coefficients are related by \begin{equation}
a_{n}=\xi_{n}\tilde{a}_{n}\,.\label{eq:a_n_vs_a_n_dach}\end{equation}
Thus, for this simple example, the transformation matrix $\mathbf{T}$
is diagonal. The matrix elements $R_{mn}=\mathrm{E}\left\{ a_{m}a_{n}^{\ast}\right\} $
and $\tilde{R}_{mn}=\mathrm{E}\left\{ \tilde{a}_{m}\tilde{a}_{n}^{\ast}\right\} $
are related by \begin{equation}
R_{mn}=\xi_{m}\tilde{R}_{mn}\xi_{n}^{\ast}\,.\label{eq:R_vs_R_dach}\end{equation}

\paragraph{Example 2 (ctd.) \cite{Kennedy2007}: }

Consider the circle (\ref{eq:Def_disk}) of radius $r_{1}$ . The
fading process $a\left(\mathbf{x}\right)$ can be written as\begin{equation}
a\left(\mathbf{x}\right)=\sum_{n=-\infty}^{\infty}\tilde{a}_{n}j^{n}\mathrm{J}_{n}\left(2\pi\left|\mathbf{x}\right|\right)\mathrm{e}^{j\beta\left(\mathbf{x}\right)n}\,,\label{eq:a(x)-1}\end{equation}
where $\beta\left(\mathbf{x}\right)$ is the polar angle of the vector
$\mathbf{x}$. This is an expansion\begin{equation}
a\left(\mathbf{x}\right)=\sum_{n=-\infty}^{\infty}\tilde{a}_{n}v_{n}\left(\mathbf{x}\right)\label{eq:Entwicklung_v_n-disk}\end{equation}
with respect to the basis\begin{equation}
v_{n}\left(\mathbf{x}\right)=j^{n}\mathrm{J}_{n}\left(2\pi\left|\mathbf{x}\right|\right)\mathrm{e}^{j\beta\left(\mathbf{x}\right)n}\,.\label{eq:v_n-disk}\end{equation}
This basis is already orthogonal. Is can be normalised by dividing
it by a normalisation factor $\xi_{n}$ that fullfills the requirement\[
\left|\xi_{n}\right|^{2}=\left\Vert v_{n}\right\Vert ^{2}=\int_{\mathcal{A}}\frac{\mathrm{d}^{2}\mathbf{x}}{\left|\mathcal{A}\right|}\,\mathrm{J}_{n}^{2}\left(2\pi\left|\mathbf{x}\right|\right)=\int_{0}^{r_{1}}\frac{2\pi r\mathrm{d}r}{\left|\mathcal{A}\right|}\,\mathrm{J}_{n}^{2}\left(2\pi r\right)\,.\]
We define\begin{equation}
\xi_{n}=j^{n}\sqrt{\int_{0}^{r_{1}}\frac{2\pi r\mathrm{d}r}{\left|\mathcal{A}\right|}\,\mathrm{J}_{n}^{2}\left(2\pi r\right)}\,.\label{eq:xi_n-disk}\end{equation}
The basis $\left\{ u_{n}\right\} _{n=-\infty}^{\infty}$ with\begin{equation}
v_{n}\left(\mathbf{x}\right)=\xi_{n}u_{n}\left(\mathbf{x}\right)\label{eq:v_n_vs_u_n-1}\end{equation}
is now given by\begin{equation}
u_{n}\left(\mathbf{x}\right)=\frac{\mathrm{J}_{n}\left(2\pi\left|\mathbf{x}\right|\right)}{\sqrt{\int_{0}^{r_{1}}\frac{2\pi r\mathrm{d}r}{\left|\mathcal{A}\right|}\,\mathrm{J}_{n}^{2}\left(2\pi r\right)}}\mathrm{e}^{j\beta\left(\mathbf{x}\right)n}\,.\label{eq:u_n-disk}\end{equation}
The coefficients are related by \begin{equation}
a_{n}=\xi_{n}\tilde{a}_{n}\,.\label{eq:a_n_vs_a_n_dach-1}\end{equation}
Thus, for this example, the transformation matrix $\mathbf{T}$ is
diagonal. The matrix elements $R_{mn}=\mathrm{E}\left\{ a_{m}a_{n}^{\ast}\right\} $
and $\tilde{R}_{mn}=\mathrm{E}\left\{ \tilde{a}_{m}\tilde{a}_{n}^{\ast}\right\} $
are related by \begin{equation}
R_{mn}=\xi_{m}\tilde{R}_{mn}\xi_{n}^{\ast}\,.\label{eq:R_vs_R_dach-1}\end{equation}

\subsection{Solving the Eigenvalue Problem}

The goal of this subsection is to approximate the eigenvalue problem
for $\mathbf{R}$ by the eigenvalue problem of a finite-dimensional
matrix. We shall replace the operators $\tilde{\mathbf{R}}$ and $\mathbf{T}$
in Equation (\ref{eq:Transf_Matrizen}) by the corresponding truncated
matrices and then use Equation (\ref{eq:eig(R)-Formel}) for these
finite-dimensional matrices to find the eigenvalues. We shall state
rigorous upper bounds for the approximation error of the corresponding
eigenvalues.

We follow the path indicated in \cite{Kennedy2007}, where the disk
aperture of Example 2 has been treated. As shown above, the transformation
matrix $\mathbf{T}$ is diagonal for that aperture. For the general
case where $\mathbf{T}$ is not diagonal, the truncation problem is
more involved. Moreover, it remains to be clarified how the truncation
to finite-rank operators influences the diversity spectrum. In the
following treatment, we shall answer these question.

\subsubsection{Truncating Errors of the Series}

As shown in \cite{Kennedy2007}, the infinite series expansion of
$a\left(\mathbf{x}\right)$ inside a disk is essentially finite-dimensional.
The authors prove that the (absolute as well as the means square)
truncation error that occurs by replacing the series (\ref{eq:Entwicklung_v_n-disk})
by the finite series\begin{equation}
a\left(\mathbf{x};N\right)=\sum_{n=-N}^{N}\tilde{a}_{n}v_{n}\left(\mathbf{x}\right)\label{eq:a_N(x)}\end{equation}
decays exponentially with $N$. These results are based on estimates
on Bessel functions that are studied in detail in Appendices I and
II of \cite{Kennedy2007}. The following theorem summarises the main
results of those appendices.

\paragraph{Theorem 1: }

Let $r_{1}\ge0$ and $N_{D}\triangleq\left\lceil \mathrm{e}\pi r_{1}\right\rceil $,
where $\left\lceil .\right\rceil $ denotes the ceiling function.
Then, for $N\ge N_{D}$ , the following two bounds hold uniformly
in $r$ for $0\le r\le r_{1}$:\begin{equation}
\sum_{\left|n\right|>N}\left|\mathrm{J}_{n}\left(2\pi r\right)\right|\le0.2\,\exp\left[N_{D}-N\right]\label{eq:Besselbound-1}\end{equation}
 \begin{equation}
\sum_{\left|n\right|>N}\mathrm{J}_{n}^{2}\left(2\pi r\right)\le0.01\,\exp\left[2\left(N_{D}-N\right)\right]\label{eq:Besselbound-2}\end{equation}

The first bound (\ref{eq:Besselbound-1}) controls the absolute truncation
error when approximating $a\left(\mathbf{x}\right)$ by $a\left(\mathbf{x};N\right)$.
Furthermore, it controls the error when $\rho\left(\mathbf{x}\right)$
given by the the series (\ref{eq:ACF-Bessel}) is approximated by
its truncated version \begin{equation}
\rho_{N}\left(\mathbf{x}\right)=\sum_{n=-N}^{N}\tilde{s}_{n}\mathrm{e}^{j\beta n}j^{n}\mathrm{J}_{n}\left(2\pi r\right)\label{eq:rho_N}\end{equation}
This latter approximation is practically important because in general
there is no closed analytical expression for $\rho\left(\mathbf{x}\right)$
available. Since the coefficients $\tilde{s}_{n}$ are bounded by
(\ref{eq:Bound-s_n}), the truncation error for $r\le r_{1}$ and
$N\ge N_{D}=\left\lceil \mathrm{e}\pi r_{1}\right\rceil $ can be
estimated as follows: \begin{eqnarray}
\left|\rho\left(\mathbf{x}\right)-\rho_{N}\left(\mathbf{x}\right)\right| & = & \left|\sum_{\left|n\right|>N}\tilde{s}_{n}\mathrm{e}^{j\beta n}j^{n}\mathrm{J}_{n}\left(2\pi r\right)\right|\nonumber \\
 & \le & \sum_{\left|n\right|>N}\left|\mathrm{J}_{n}\left(2\pi r\right)\right|\nonumber \\
 & \le & 0.2\,\mathrm{e}^{N_{D}-N}\label{eq:trunc-error-rho}\end{eqnarray}

The second bound (\ref{eq:Besselbound-2}) controls the means square
(MS) truncation error of the series expansions for $a\left(\mathbf{x}\right)$.
For Examples 1 and 2, the MS truncation error of the series (\ref{eq:Entwicklung_v_n})
and (\ref{eq:Entwicklung_v_n-disk}) can be estimated as follows:\begin{eqnarray}
\mathrm{E}\left\{ \left\Vert \sum_{\left|n\right|>N}\tilde{a}_{n}v_{n}\right\Vert ^{2}\right\}  & = & \mathrm{E}\left\{ \left\Vert \sum_{\left|n\right|>N}\tilde{a}_{n}\xi_{n}u_{n}\right\Vert ^{2}\right\} \nonumber \\
 & = & \sum_{\left|n\right|>N}\mathrm{E}\left\{ \left|\tilde{a}_{n}\right|^{2}\right\} \left|\xi_{n}\right|^{2}\nonumber \\
 & = & \sum_{\left|n\right|>N}\left|\xi_{n}\right|^{2}\label{eq:truncation_error-1}\end{eqnarray}
Here we have used $v_{n}=\xi_{n}u_{n}$, the orthonormality of the
base $u_{n}$, and the property \[
\mathrm{E}\left\{ \left|\tilde{a}_{n}\right|^{2}\right\} =\tilde{s}_{0}=1\,\]
that follows from Equation (\ref{eq:R_mn_tilde_FK}). In both examples,
the coefficients $\xi_{n}$ can be expressed by Bessel functions,
see Equation (\ref{eq:xi_n-circle}) for the circle and Equation (\ref{eq:xi_n-disk})
for the disk. For the circle, $\left|\xi_{n}\right|^{2}=\mathrm{J}_{n}^{2}\left(2\pi r\right)$
holds, and we obtain the truncation error\begin{eqnarray*}
\mathrm{E}\left\{ \left\Vert \sum_{\left|n\right|>N}\tilde{a}_{n}v_{n}\right\Vert ^{2}\right\}  & \le & \sum_{\left|n\right|>N}\mathrm{J}_{n}^{2}\left(2\pi r\right)\\
 & \le & 0.01\,\exp\left[2\left(N_{D}-N\right)\right]\end{eqnarray*}
with $N_{D}\triangleq\left\lceil \mathrm{e}\pi r\right\rceil \,.$
For the disk, we insert the expression (\ref{eq:xi_n-disk}) for $\xi_{n}$
and obtain

\begin{eqnarray*}
\mathrm{E}\left\{ \left\Vert \sum_{\left|n\right|>N}\tilde{a}_{n}v_{n}\right\Vert ^{2}\right\}  & = & \sum_{\left|n\right|>N}\left|\xi_{n}\right|^{2}\\
 & = & \int_{0}^{r_{1}}\frac{2\pi}{\left|\mathcal{A}\right|}r\mathrm{d}r\,\sum_{\left|n\right|>N}\mathrm{J}_{n}^{2}\left(2\pi r\right)\\
 & \le & 0.01\,\exp\left[2\left(N_{D}-N\right)\right]\int_{0}^{r_{1}}\frac{2\pi}{\left|\mathcal{A}\right|}r\mathrm{d}r\,\end{eqnarray*}
with $N_{D}\triangleq\left\lceil \mathrm{e}\pi r_{1}\right\rceil $
. The integral equals one and we obtain the same bound as for the
circle of the same radius:\begin{equation}
\mathrm{E}\left\{ \left\Vert \sum_{\left|n\right|>N}\tilde{a}_{n}v_{n}\right\Vert ^{2}\right\} \le0.01\,\exp\left[2\left(N_{D}-N\right)\right]\,\label{eq:Trunc_Error_circle_disk}\end{equation}
This is just the statement of Theorem 2 in \cite{Kennedy2007}.

\subsubsection{Approximations for Operators and Eigenvalues}

For the general case where the transformation matrix $\mathbf{T}$
is not diagonal, the formalism is more involved than in the two examples
discussed above. Before going into details, we explain what problem
has to be solved.

Let $\mathbf{R}_{N}$ be the truncated version of the inifinite autocorrelation
matrix $\mathbf{R}$ in which all elements $R_{mn}$ with $\left|m\right|>N$
and $\left|n\right|>N$ are set to zero. Heuristically, it is evident
that $\mathbf{R}_{N}\rightarrow\mathbf{R}$ in some sense as $N\rightarrow\infty$.
The appropriate type of convergence is specified and proven in Appendices
A and B. From $\mathbf{R}_{N}\rightarrow\mathbf{R}$ we then obtain
convergence of the corresponding eigenvalue spectra.

We write $\lambda_{i}\left(\mathbf{R}_{N}\right)$ and $\lambda_{i}\left(\mathbf{R}\right)$
for the eigenvalue number $i$ of $\mathbf{R}_{N}$ and $\mathbf{R}$,
respectively. Assume that both eigenvalue spectra are numbered in
descending order:\begin{equation}
\lambda_{1}\ge\lambda_{2}\ge\lambda_{3}\ge...\label{eq:eigenvalue_ordering}\end{equation}
Recall that, since both operators are compact, they have discrete
spectra.  We shall have to prove that $\lambda_{i}\left(\mathbf{R}_{N}\right)\to\lambda_{i}\left(\mathbf{R}\right)$
 holds for $\ensuremath{N\to\infty}$. Based on the estimates from
\cite{Kennedy2007} as stated in Theorem 1, we are able to prove the
following

\paragraph{Theorem 2:}

Let $\mathcal{A}\subset\mathcal{D}$ where $\mathcal{D}$ is a disk
of radius $r_{1}$. Assume that $\mathcal{S}\left(\alpha\right)$
is a piecewise continuous, bounded function over the interval $\left[-\pi,\pi\right]$.
Then the inequality \begin{equation}
\left|\lambda_{i}\left(\mathbf{R}\right)-\lambda_{i}\left(\mathbf{R}_{N}\right)\right|\le0.2\rho_{max}\exp\left(N_{D}-N\right)\,\label{eq:Eigenwert-Schranke}\end{equation}
holds for $N\ge N_{D}$ and $N_{D}=\left\lceil \mathrm{e}\pi r_{1}\right\rceil $
and with a constant $\rho_{max}=\max_{\alpha\in\left[-\pi,\pi\right]}\left(2\pi\mathcal{S}\left(\alpha\right)\right)$.

The proof is given in Appendix B.

Now let $\mathbf{T}_{N}$ and $\mathbf{G}_{N}=\mathbf{T}_{N}^{\dagger}\mathbf{T}_{N}$
be the truncated versions of $\mathbf{T}$ and $\mathbf{G}$, respectively.
Then $\mathbf{R}_{N}=\mathbf{T}_{N}\tilde{\mathbf{R}}\mathbf{T}_{N}^{\dagger}$,
and the approximate eigenvalues can be numerically calculated from\textbf{
\begin{equation}
\mathrm{eig}\left(\mathbf{R}_{N}\right)=\mathrm{eig}\left(\tilde{\mathbf{R}}\mathbf{G}_{N}\right)\,.\label{eq:eig(R_N)-Formel}\end{equation}
}Examples of such numerically calculated eigenvalue spectra will be
shown in the next section.

According to Theorem 2, the eigenvalues of $\mathbf{R}$ can be approximated
exponentially tight by the eigenvalues of $\mathbf{R}_{N}$. One might
expect that this has the consequence that the diversity measure $\omega\left(\mathbf{R}\right)$
given by Equation (\ref{eq:Diversity_measure-4}) can be approximated
tightly by \begin{equation}
\omega\left(\mathbf{R}_{N}\right)=\frac{1}{\bigl\Vert\mathbf{R}_{N}\bigr\Vert_{2}^{2}}\,.\label{eq:omega_N}\end{equation}
In fact, as shown in Appendix 3, the following tight approximation
holds:

\paragraph{Theorem 3:}

Let $\mathcal{A}\subset\mathcal{D}$ where $\mathcal{D}$ is a disk
of radius $r_{1}$. Assume that $\mathcal{S}\left(\alpha\right)$
is a piecewise continuous, bounded function over the interval $\left[-\pi,\pi\right]$.
Then the inequality \begin{equation}
\left|\bigl\Vert\mathbf{R}\bigr\Vert_{2}^{2}-\bigl\Vert\mathbf{R}_{N}\bigr\Vert_{2}^{2}\right|\le0.4\rho_{max}^{2}\exp\left(N_{D}-N\right)\,\label{eq: HS-Norm-Schranke}\end{equation}
holds for $N\ge N_{D}$ and $N_{D}=\left\lceil \mathrm{e}\pi r_{1}\right\rceil $
and with a constant $\rho_{max}=\max_{\alpha\in\left[-\pi,\pi\right]}\left(2\pi\mathcal{S}\left(\alpha\right)\right)$.

The statement of the theorem means that the \emph{inverse} diversity
measure $\omega^{-1}\left(\mathbf{R}\right)=\bigl\Vert\mathbf{R}\bigr\Vert_{2}^{2}$
can be approximated exponentially tight by the corresponding truncated
quantity $\omega^{-1}\left(\mathbf{R}_{N}\right)=\bigl\Vert\mathbf{R}_{N}\bigr\Vert_{2}^{2}$
. These quantities are of primary relevance, because for the calculation
of the slope $\Sigma_{0}$ of a MIMO system according to Equation
(\ref{eq:Slope-Formel}), these inverse diversity measures for the
transmit and receive apertures have to be added. The approximation
error of a numerically calculated diversity measure $\omega\left(\mathbf{R}_{N}\right)$
itself can be estimated from the geometric series argument \begin{equation}
\omega\left(\mathbf{R}\right)=\omega\left(\mathbf{R}_{N}\right)\frac{1}{1-\epsilon}=\omega\left(\mathbf{R}_{N}\right)\left(1+\epsilon+\mathcal{O}\left(\epsilon^{2}\right)\right)\;\label{eq:geom-series-argument}\end{equation}
with an error $\epsilon=\omega\left(\mathbf{R}_{N}\right)\left(\bigl\Vert\mathbf{R}_{N}\bigr\Vert_{2}^{2}-\bigl\Vert\mathbf{R}\bigr\Vert_{2}^{2}\right)$
that can be controlled by Equation (\ref{eq: HS-Norm-Schranke}).

\subsection{A Generalisation: $M$ parallel Lines }

\begin{figure}
\noindent \begin{centering}
\includegraphics[scale=1.2]{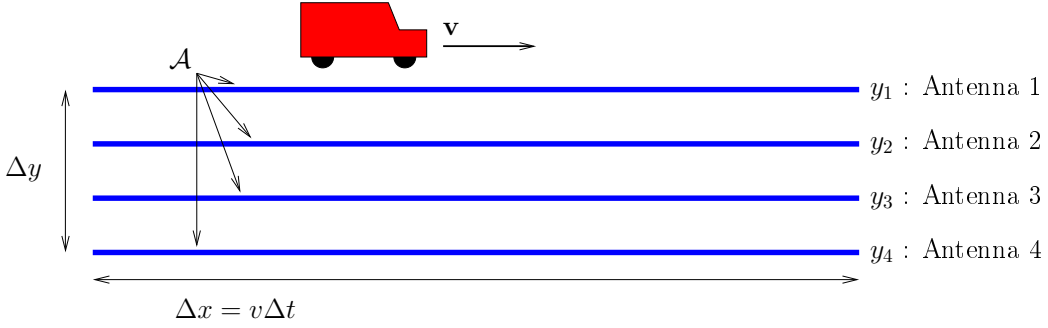}
\par\end{centering}

\caption{\label{cap:vehicle-4antennas} Four antennas mounted on a vehicle
moving with velocity $v$. }

\end{figure}

Let us consider $M$ antennas mounted transverse across the roof of
a vehicle that is moving into x-direction, see Figure \ref{cap:vehicle-4antennas}.
Then the aperture $\mathcal{A}$ consists of $M$ parallel horizontal
lines. Time diversity provided by time interleaving corresponds to
the x-direction, antenna diversity to the y-direction.%

Because this aperture is not a continuous curve, the results above
can not applied directly. However, there is a simple argument how
they can be adopted to this important case. 

The problem can be solved by constructing a Hilbert space $\mathcal{H}$
on that aperture that consists of $M$ straight lines. The construction
is just a slightly modified direct sum \cite{Reed-Simon-1} of the
$M$ Hilbert spaces $\mathcal{H}_{m}$, $m=1,...,M$ corresponding
to these $M$ straigth lines. Let $\phi_{m},\,\psi_{m}\in\mathcal{H}_{m}$.
Their scalar product is denoted as $\left\langle \phi,\psi\right\rangle _{m}$.
The vectors $\phi,\,\psi\in\mathcal{H}$ are then defined as $M$-tuples
$\phi=\left(\phi_{1},...,\phi_{M}\right)^{T}$, $\psi=\left(\psi_{1},...,\psi_{M}\right)^{T}$
that, for convenience, may be written as columns. Their scalar product
is defined by\begin{equation}
\left\langle \phi,\psi\right\rangle =\frac{1}{M}\sum_{m=1}^{M}\left\langle \phi_{m},\psi_{m}\right\rangle \,.\label{eq:Skalarprodukt-direkte-Summe}\end{equation}
Let $\tau$ be the common priviledged parameter of all the lines.
Then the scalar product can be written as \begin{equation}
\left\langle \phi,\psi\right\rangle =\frac{1}{M}\sum_{m=1}^{M}\int_{0}^{1}\phi_{m}^{\ast}\left(\tau\right)\psi\left(\tau\right)\mathrm{d}\tau\,.\label{eq:Skalarprodukt-direkte-Summe-1}\end{equation}
The fading process itself is a column vector $a=\left(a_{1},...,a_{M}\right)^{T}$.
The corresponding autocorrelation operator $\mathbf{R}$ will then
be given according to the general definition (\ref{eq:Def_R-1}) as
\begin{equation}
\left\langle \phi,\mathbf{R}\psi\right\rangle =\mathrm{E}\left\{ \left\langle \phi,a\right\rangle \left\langle a,\psi\right\rangle \right\} \,.\label{eq:Def_R-1-direkte_summe}\end{equation}
The corresponding integral kernel now becomes a tensor $\mathcal{R}_{mm^{\prime}}\left(\tau,\tau^{\prime}\right)$,
$m,m^{\prime}\in\left\{ 1,...,M\right\} $. This kernel is a continuous
matrix-valued function in $\tau$ and $\tau^{\prime}$. As a consequence
of that continuity, all the results above remain valid and our method
can be applied to this case.

\section{Numerically calculated diversity spectra}

\subsection{How to Calculate Diversity Spectra }

The task now is to calculate the eigenvalue spectrum according to
Equation (\ref{eq:eig(R_N)-Formel}). It depends on the geometry of
$\mathcal{A}$ and on the Fourier coefficients $\hat{s}_{n}$ of the
PAS $\mathcal{S}\left(\alpha\right)$. The elements of $\mathbf{G}_{N}$
are given by $G_{mn}=\left\langle v_{m},v_{n}\right\rangle $. This
scalar product is an integral over $\mathcal{A}$ and has to be calculated
by numerical quadrature methods for all $m,\, n$ with $\left|m\right|,\,\left|n\right|\le N$.
The matrix elements of $\tilde{\mathbf{R}}$ are given by $\tilde{R}_{mn}=2\pi\hat{s}_{m-n}$.
The truncation number $N$ governs the accuracy of the eigenvalues
according to Theorem 2. For the choise $N=N_{D}+10$, e.g., the exponential
is approximately $5\cdot10^{-5}$, which can be regarded as sufficiently
accurate. It is noteworthy that the constant $0.2\cdot\max_{\alpha\in\left[-\pi,\pi\right]}\left(2\pi\mathcal{S}\left(\alpha\right)\right)$
in that theorem is typically in the order of one%
\footnote{For the isotropical PAS $\mathcal{S}\left(\alpha\right)=\left(2\pi\right)^{-1}$,
as an example, it has the value 0.2. For a uniform distribution $\mathcal{S}\left(\alpha\right)=\Delta^{-1}rect\left(\alpha/\Delta\right)$
over an opening angle $\Delta$, it has the value $0.2\cdot2\pi/\Delta$,
which does not exceed the order of one unless the opening angle is
extremely narrow. 

}. The number $N_{D}=\left\lceil \mathrm{e}\pi r_{1}\right\rceil $
grows with the size of the aperture $\mathcal{A}$. If, e.g., it is
inside a disk with radius of one wavelength, $N_{D}=\left\lceil \mathrm{e}\pi\right\rceil =9$.
Thus, for typical aperture restricted to the size of a few wavelengths,
the matrices do not exceed the size of a few times ten.

\subsection{The Power Azimuth Spectra}

We consider two PAS prototypes that are quite popular in the literature:
 The uniform PAS and the von-Mises PAS.

\paragraph{Uniform PAS:}

By this we mean a PAS that is constant over an opening angle $\Delta$.
The normalised and centered uniform PAS of opening angle $\Delta$
is given by \begin{equation}
\mathcal{S}\left(\alpha;\Delta\right)=\frac{1}{\Delta}\mathrm{rect}\left(\frac{\alpha}{\Delta}\right)\,.\label{eq:PAS-uniform}\end{equation}
The corresponding Fourier coefficients can be calculated as\begin{equation}
\hat{s}_{n}=\frac{1}{2\pi}\mathrm{sinc}\left(n\frac{\Delta}{2\pi}\right)\,.\label{eq:FK-uniform}\end{equation}

\paragraph{von-Mises PAS:}

The von-Mises distribution is a circular distribution with a shape
quite similar to a wrapped normal distribution \cite{Ravindran}.
Its shape is controlled by a parameter $\kappa\ge0$, whose inverse
is approximately equal to the variance of the corresponding normal
distribution. The centered PAS is given by \begin{equation}
\mathcal{S}\left(\alpha;\kappa\right)=\frac{1}{2\pi\mathrm{I}_{0}\left(\kappa\right)}\mathrm{e}^{\kappa\cos\alpha}\,,\label{eq:PAS-vM}\end{equation}
where $\mathrm{I}_{n}\left(\kappa\right)$ denotes the modified Bessel
function. The Fourier coefficients\begin{equation}
\hat{s}_{n}=\frac{\mathrm{I}_{n}\left(\kappa\right)}{2\pi\mathrm{I}_{0}\left(\kappa\right)}\label{eq:FK-vM}\end{equation}
can be obtain from Formula 9.6.34 in \cite{AbramowitzStegun}.

The formulas above apply to a PAS that is centered around $\alpha=0$.
The versions centered around $\alpha=\alpha_{0}$ are obtained by
the shift \begin{equation}
\mathcal{S}\left(\alpha\right)\mapsto\mathcal{S}\left(\alpha-\alpha_{0}\right)\,.\label{eq:-2}\end{equation}
The corresponding Fourier coefficients must then be modified according
to\[
\hat{s}_{n}\mapsto\hat{s}_{n}\mathrm{e}^{-j\alpha_{0}n}\,.\]

\subsection{Uniform Circular Arrays}

\begin{figure}
\noindent \begin{centering}
\includegraphics[scale=0.7]{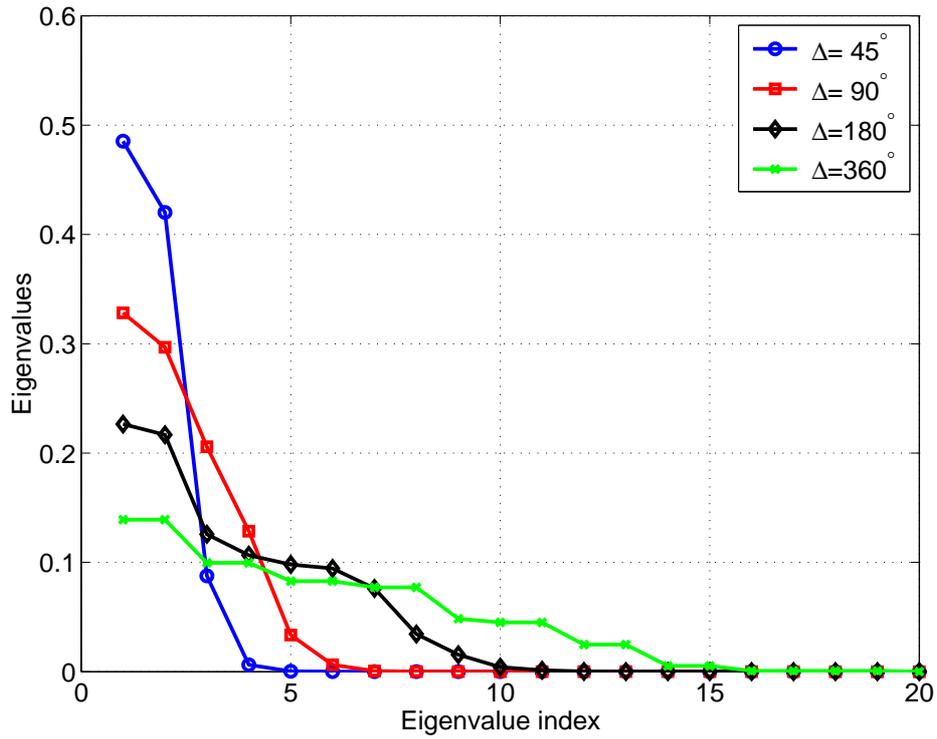}
\par\end{centering}

\caption{\label{cap:EW-UCA-uni-r=00003D1} Diversity spectra for a dense UCA
of radius $r=1$ for a uniform PAS with different opening angles $\Delta$. }

\end{figure}

\begin{figure}
\noindent \begin{centering}
\includegraphics[scale=0.7]{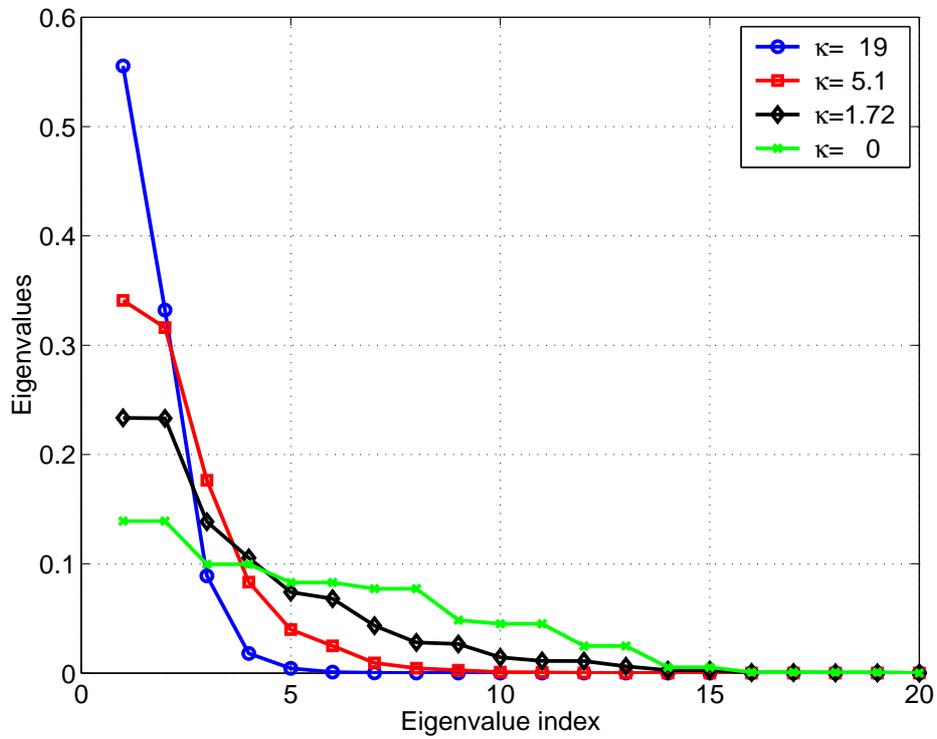}
\par\end{centering}

\caption{\label{cap:EW-UCA-vM-r=00003D1} Diversity spectra for a dense UCA
of radius $r=1$ for a von-Mises PAS with different parameters $\kappa$. }

\end{figure}

A uniform circular array (UCA) is a configuration where the antennas
are mounted uniformly on a circle. In the limit of a dense array,
this is just the situation of Example 1 discussed above. Figures \ref{cap:EW-UCA-uni-r=00003D1}
and \ref{cap:EW-UCA-vM-r=00003D1} show the numerically calculated
eigenvalue spectra for a ULA of radius $r=1$ (i. e. one wavelenght)
for the uniform and the von-Mises PAS, respectively. The parameters
$\Delta$ and $\kappa$ are adjusted manually in such a way that the
corresponding values for $\omega$ are nearly identical. The eigenvalue
spectra are very similar, but, however, they are not identical. Obviously,
the multipath richness decreases significantly for small opening angles
of the PAS, i.e. for small values of $\Delta$ and for high values
of $\kappa$.  %

\begin{figure}
\noindent \begin{centering}
\includegraphics[scale=0.7]{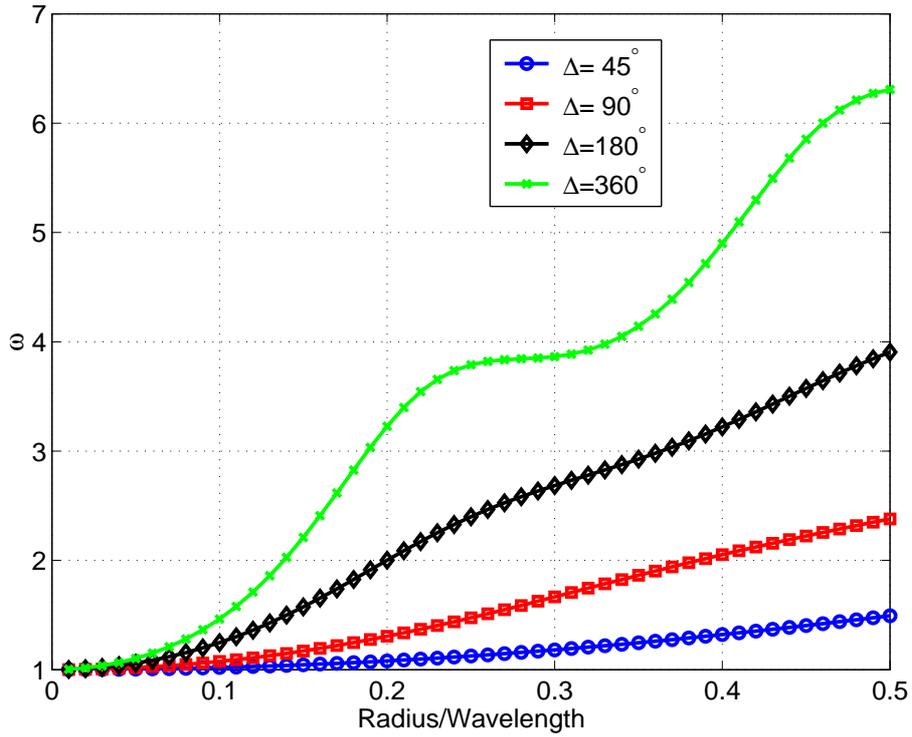}
\par\end{centering}

\caption{\label{cap:UCA_PAS=00003Duniform_2010-9-4-15-15} The diversity measure
$\omega$ as a function of the radius for a dense UCA and a uniform
PAS with different opening angles $\Delta$. }

\end{figure}

\begin{figure}
\noindent \begin{centering}
\includegraphics[scale=0.7]{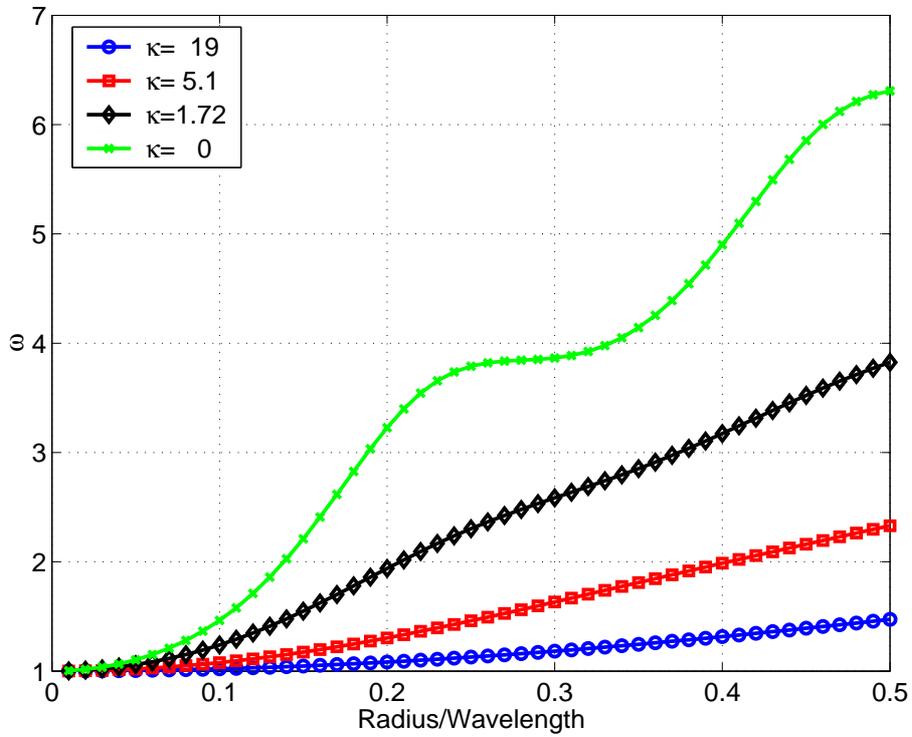}
\par\end{centering}

\caption{\label{cap:UCA_PAS=00003Dvon-Mises_2010-9-4-15-15} The diversity
measure $\omega$ as a function of the radius for a dense UCA and
a von-Mises PAS with different parameters $\kappa$. }

\end{figure}

Figures \ref{cap:UCA_PAS=00003Duniform_2010-9-4-15-15} and \ref{cap:UCA_PAS=00003Dvon-Mises_2010-9-4-15-15}
show $\omega$ %
 for small radii up to $r=0.5$ for the same parameters of $\Delta$
and $\kappa$. The curves are nearly identical in both figures. The
green curve of the isotropic PAS ($\Delta=360^{\circ}$ or $\kappa=0$)
shows a noticable oscillatory behaviour, that can also be observed
for other values that are close to the isotropic case.

\subsection{Uniform Linear Arrays}

For a straigth line which corresponds to a uniform linear array (ULA),%
\begin{figure}
\noindent \begin{centering}
\includegraphics[scale=0.7]{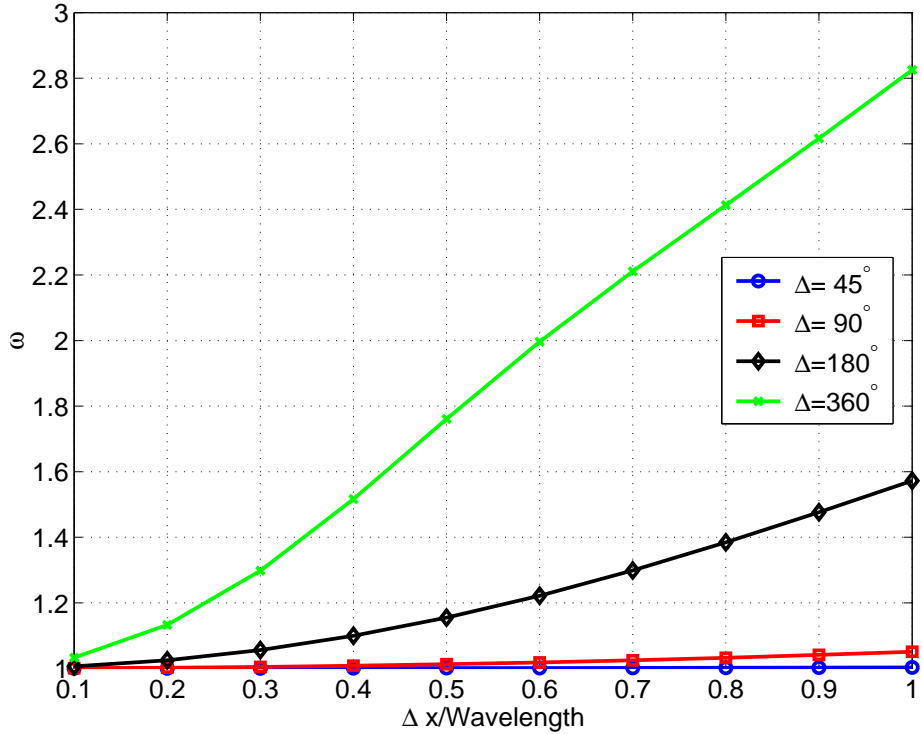}
\par\end{centering}

\caption{\label{cap:Diversity_Spectra/Figures/ULA_PAS=00003Duniform_richtung=00003D0_2010-9-5-12-39}
The diversity measure $\omega$ as a function of the size $\Delta x$
for a dense ULA and a uniform PAS with different opening angles $\Delta$.
The main direction of arrival is $\alpha_{0}=0$, i.e. parallel to
the array.}

\end{figure}
 such an oscillatory behaviour could not be observed. Figure \ref{cap:Diversity_Spectra/Figures/ULA_PAS=00003Duniform_richtung=00003D0_2010-9-5-12-39}
shows $\omega$ for ULA sizes up to one wavelenght and the uniform
PAS. The direction of the ULA is parallel to the mean angle of arrial
($\alpha_{0}=0$). This is the worst case, and for small values of
$\Delta$ (blue and red curve), there is practically no diversity
gain. Increasing the size of the ULA up to several wavelengths does
not really help because the steepness of the curves is very poor.
There is a strong dependency on the mean angle of arrival $\alpha_{0}$.
Figure \ref{cap:ULA_PAS=00003Duniform_laenge=00003D10_2010-9-6-17-2}
\begin{figure}
\noindent \begin{centering}
\includegraphics[scale=0.7]{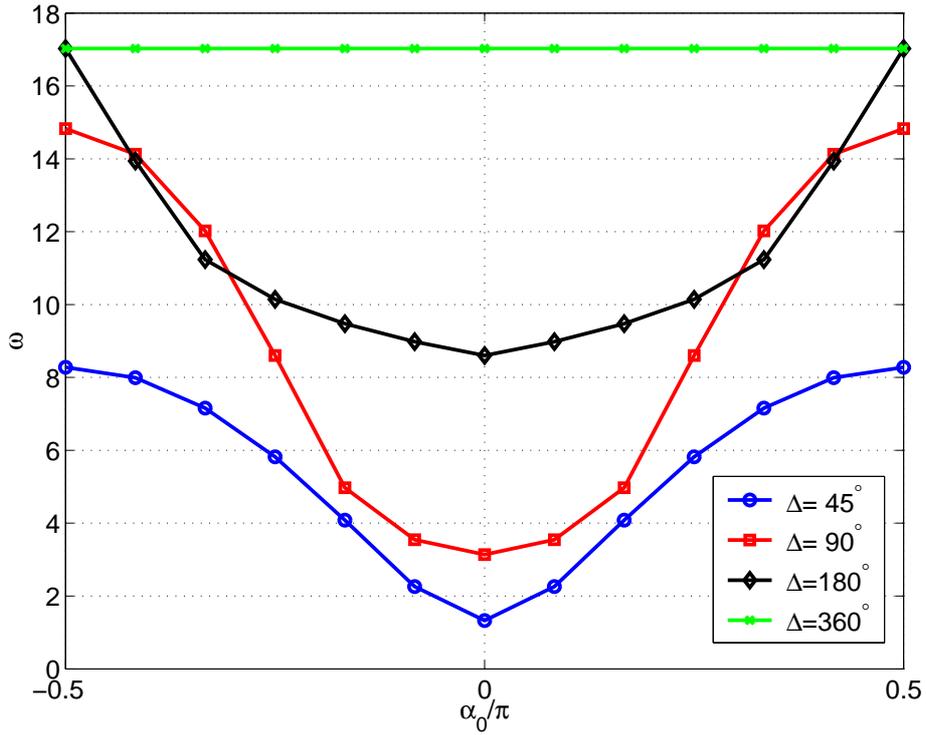}
\par\end{centering}

\caption{\label{cap:ULA_PAS=00003Duniform_laenge=00003D10_2010-9-6-17-2} The
diversity measure $\omega$ as a function of the main direction of
arrival $\alpha_{0}$ for a dense ULA and a uniform PAS with different
opening angles $\Delta$. The size of the array is $\Delta x=10$.}

\end{figure}
 shows $\omega$ as a function of $\alpha_{0}$ for $\Delta x=10$.
It can be seen from this figure that even for this large array, the
diversity measure falls down close to one for small opening angles
(blue and red curve). This means that for a narrowly focussed signal,
even a huge antenna array parallel to the direction of arrival can
hardly provide any diversity gain.  Such narrowly focussed signal
may be not very likely at the mobile station, but they will frequently
occur at the base station site. Therefore it is important at that
site not to fix all antennas along one line, but to distribute them
in two dimensions by using, e.g., a circular array or any other two-dimensional
configuration. Figure \ref{cap:DABDIVDOC/Diversity_Spectra/Figures/ULA_PAS=00003Duniform_laenge=00003D1_2010-9-6-16-7}%
\begin{figure}
\noindent \begin{centering}
\includegraphics[scale=0.7]{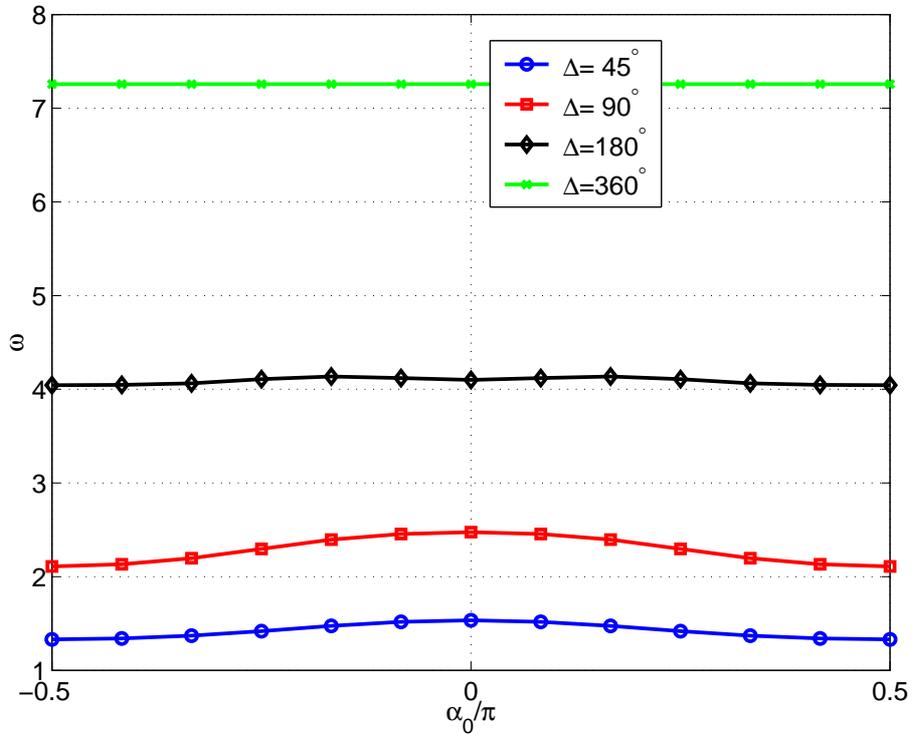}
\par\end{centering}

\caption{\label{cap:DABDIVDOC/Diversity_Spectra/Figures/ULA_PAS=00003Duniform_laenge=00003D1_2010-9-6-16-7}
The diversity measure $\omega$ as a function of the main direction
of arrival $\alpha_{0}$ for four parallel and equidistant lines and
a uniform PAS with different opening angles $\Delta$. The length
of the lines is $\Delta x=1$ and the separation of the outer lines
is $\Delta y=1$.}

\end{figure}
 shows $\omega$ as a function of $\alpha_{0}$ for a configuration
of four parallel ULAs of length $\Delta x=1$ and a separation of
the outer ULAs given by $\Delta y=1$. In that scenario, the diversity
measure is very robust against rotations. As discussed above (see
Figure \ref{cap:vehicle-4antennas}), by space-time scaling $\Delta x=v\Delta t$,
this situation can also be related to the scenario of a vehicle moving
in x-direction with velocity $v$ and a 4-antenna communication system
with time interleaving over a interval $\Delta t$.

\subsection{Discrete Arrays}

The focus of this paper is to analyse the diversity spectra of continuous
(dense) arrays, i.e. of contiuous regions in the plane. It is interesting
to compare this with the discrete case where a finite number $L$
of antennas is mounted on a finite geometrical base, i.e. a circle
for the UCA or a line for the ULA. In \cite{Muharemovic2008}, a detailed
analysis has been presented how the diversity measure $\omega$ depends
on the array geometry. That analysis, however, is restricted to ideal
case of an isotropic PAS where the autocorrelation function is given
by $\rho_{0}\left(\mathbf{x}\right)=\mathrm{J}_{0}\left(2\pi r\right)$.
In the general case, the ACF is given by the expansion (\ref{eq:ACF-Bessel}).
According to Theorem 1, this can be approximated by the truncated
series\begin{equation}
\rho_{N}\left(\mathbf{x}\right)=\sum_{n=-N}^{N}\tilde{s}_{n}\mathrm{e}^{j\beta n}j^{n}\mathrm{J}_{n}\left(2\pi r\right)\,.\label{eq:rho_N-1}\end{equation}
The appropriate order $N$ can be chosen as $N=N_{D}+10$ according
to the bounds obtained in the last section. The diversity measure
$\omega$ according to Equation (\ref{eq:Diversity_measure-3}) has
be calculated from the matrix $\mathbf{R}$ with elements \begin{equation}
R_{ik}=\rho\left(\mathbf{x}_{i}-\mathbf{x}_{k}\right)\approx\rho_{N}\left(\mathbf{x}_{i}-\mathbf{x}_{k}\right)\,,\label{eq:R_ik-diskret}\end{equation}
where $\mathbf{x}_{i}$, $i=1,...,L$ are the antenna positions. We
note that Equation (\ref{eq:Diversity_measure-3}) does not require
to solve the eigenvalue problem for $\mathbf{R}$. Only the traces
of the matrices $\mathbf{R}$ and $\mathbf{R}^{\dagger}\mathbf{R}$
need to be calculated. 

\begin{figure}
\noindent \begin{centering}
\includegraphics[scale=0.7]{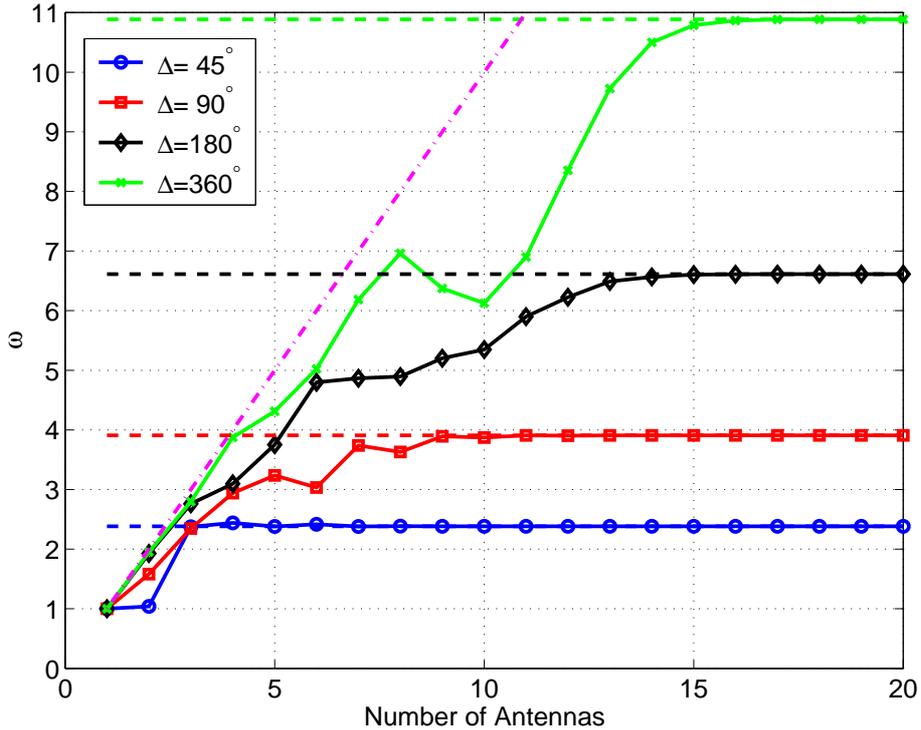}
\par\end{centering}

\caption{\label{cap:UCAdiskret_PAS=00003Duniform_r=00003D1-2010-9-7-9-10}
The diversity measure $\omega$ as a function of the number of antennas
for a UCA of radius $r=1$ and a uniform PAS with different opening
angles $\Delta$. }

\end{figure}

Figure \ref{cap:UCAdiskret_PAS=00003Duniform_r=00003D1-2010-9-7-9-10}
 shows the depency of $\omega$ on the number $L$ of antennas for
a UCA of radius $r=1$ and a uniform PAS of different opening angles
$\Delta$. The margenta dash-dotted line corresponds to $\omega=L$
(i.e. the diversity measure of $L$ uncorrelated antennas). The horizontal
dashed lines mark the $\omega$- values for the continuous UCA and
may be interpreted as the asymptotic limit $L\to\infty$. It can be
seen from the figure that for large $L$, the diversity measure runs
into a saturation given by that continuous case asymptote. It is interesting
to note that $\omega$ does not increase monotonically with $L$.
This interesting fact has been already observed in \cite{Muharemovic2008}
for the isotropic case (i.e. $\Delta=360^{\circ}$)%
\footnote{The green curve of Figure \ref{cap:UCAdiskret_PAS=00003Duniform_r=00003D1-2010-9-7-9-10}
is identical to the black curve of Figure 4 in that paper.%
}. In the same paper, it has also been pointed out that for a ULA,
the maximal value of $\omega$ is typically reached for a relatively
small number of antennas, and the limiting value for $L\to\infty$
lies below that value. %
\begin{figure}
\noindent \begin{centering}
\includegraphics[scale=0.7]{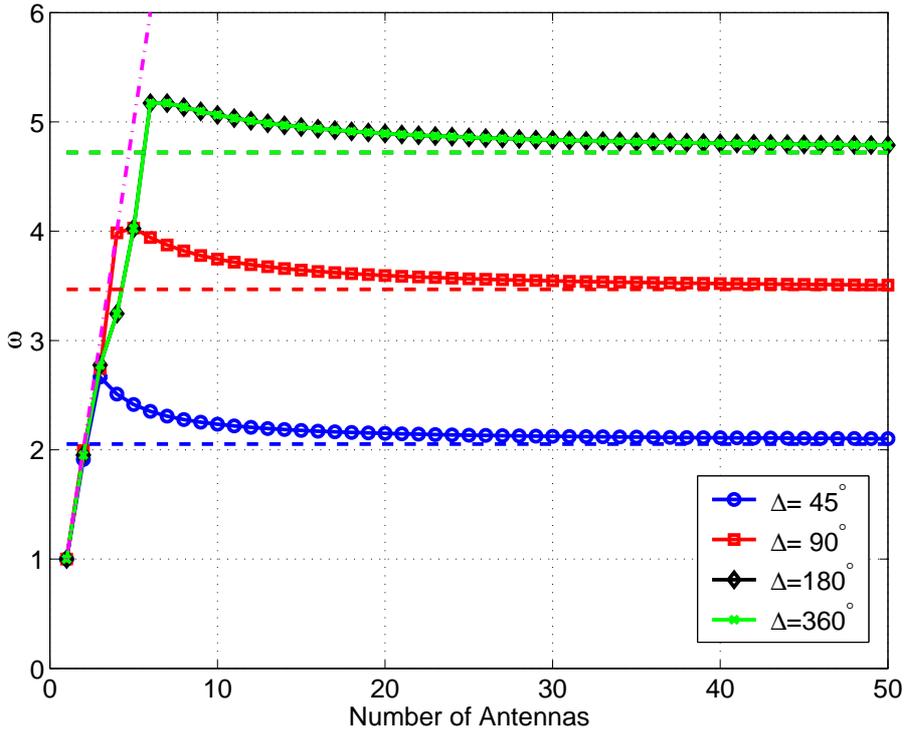}
\par\end{centering}

\caption{\label{cap:ULAdiskret_PAS=00003Duniform_x=00003D2Richtung=00003D0.5-2010-9-7-11-59}The
diversity measure $\omega$ as a function of the number of antennas
for a ULA of size $\Delta x=2$ and a uniform PAS with different opening
angles $\Delta$. The signal comes from broadside ($\alpha_{0}=\pi/2$).}

\end{figure}
 Figure \ref{cap:ULAdiskret_PAS=00003Duniform_x=00003D2Richtung=00003D0.5-2010-9-7-11-59}
shows the situation for a ULA of size $\Delta x=2$, $\alpha_{0}=90^{\circ}$
and the same PAS parameters as above. For $\Delta=360^{\circ}$, the
maximal value of $\omega$ is reached at the number of $L=6$ antenna.
For $\Delta=90^{\circ}$ and $\Delta=45^{\circ}$, this number is
even lower. The black curve for $\Delta=180^{\circ}$ hidden by the
green curve corresponding to $\Delta=360^{\circ}$ because of the
mirror symmetry of that geometrical configuration as already discussed
in the context of the Doppler effect. Figure \ref{cap:ULAdiskret_PAS=00003Duniform_x=00003D2Richtung=00003D0.5-2010-9-7-11-59}
(or similar figures for other values of $\Delta x$ ) provides practical
design guidance for the design of linear arrays. The number of antennas
should be chosen in such a way that $\omega$ is not only maximal
for a special case like $\Delta=360^{\circ}$ but should stay robust
against a change of the PAS.

\section{Discussion and Conclusions}

The diversity corresponding to an aperture $\mathcal{A}$ is characterised
by the diversity spectrum $\left\{ \lambda\right\} _{i=1}^{\infty}$
of eigenvalues of the spatial autocorrelation operator $\mathbf{R}$
that acts as an integral kernel on the Hilbert space of square integrable
functions defined on that aperture. The diversity spectrum depends
on the geometry of the aperture and the power azimuth spectrum that
characterises the statistics of the spatial fading. We have shown
a method for the numerical calculation solving the eigenvalue problem
for an finite-dimension approximation of $\mathbf{R}$. The method
provides rigorous bounds of the approximation error. A simpler, but
less comprehensive characterisation compared to the whole diversity
spectrum $\mathrm{eig}\left(\mathbf{R}\right)=\left\{ \lambda\right\} _{i=1}^{\infty}$
is a single number named the diversity measure $\omega\left(\mathbf{R}\right)$.
It can (roughly) be interpreted as the effective number of equivent
independent diversity branches. Its calculation is in the same framework,
but it is simpler because no eigenvalue problem has to be solved.
The dependency of $\omega\left(\mathbf{R}\right)$ on the aperture
geometry, on the main angle of incidence, and on the anisotropy of
the PAS gives useful hints for the design of antenna arrays. For a
vehicle of constant velocity moving through the spatial fading pattern,
the time dependent fading can be obtained by a simple scaling, and
a time interval corresponds to a spatial line segment. Therefore,
the time interleaving depth for a coded system can be assigned to
a line segment, and the diversity of that segment corresponds to the
code diversity that can be exploited within the time interleaving
depth.

\appendix

\section{Some Definitions and Facts about linear Operators}

Let $\mathbf{A}$ be a linear operator on $\mathcal{H}$. Its (hermitian)
\emph{adjoint, $\mathbf{A}^{\dagger}$, }is defined by \begin{equation}
\left\langle \phi,\mathbf{A}^{\dagger}\psi\right\rangle =\left\langle \mathbf{A}\phi,\psi\right\rangle \,.\label{eq:Def_Adjoint}\end{equation}
If $\mathbf{A}=\mathbf{A}^{\dagger}$ holds, the operator is called
\emph{self-adjoint}. In this paper, we need only to consider bounded
operators. For these operator, self-adjoint is the same as the (in
general weaker) property \emph{symmetric} or \emph{hermitian}. 

A limit $\mathbf{R}_{N}\rightarrow\mathbf{R}$ for operators may be
defined in different meanings. In this paper, we need three types
of norms for an operator $\mathbf{A}$ on a Hilbert space $\mathcal{H}$.
We briefly recall their definitions and their most important properties
and refer to \cite{Reed-Simon-1} for further details. 
\begin{enumerate}
\item The (\emph{uniform})\emph{ operator norm} $\left\Vert \mathbf{A}\right\Vert $
is defined by\begin{equation}
\left\Vert \mathbf{A}\right\Vert \triangleq\sup_{\left\Vert \phi\right\Vert =1}\left\Vert \mathbf{A}\phi\right\Vert \,.\label{eq:Uniform_Norm}\end{equation}
Operators with $\left\Vert \mathbf{A}\right\Vert <\infty$ are called
bounded operators. 
\item The \emph{Hilbert-Schmidt norm }$\left\Vert \mathbf{A}\right\Vert _{2}$
is defined as \begin{equation}
\left\Vert \mathbf{A}\right\Vert _{2}\triangleq\sqrt{\mathrm{tr}\left(\mathbf{A}^{\dagger}\mathbf{A}\right)}\,,\label{eq:HS-Norm}\end{equation}
where $\mathrm{tr}\left(\cdot\right)$ denotes the trace of an operator.
For the special case of finite-dimensional matrices, the Hilbert-Schmidt
norm is called \emph{Frobenius }norm. One can show that \begin{equation}
\left\Vert \mathbf{A}\right\Vert _{2}=\sqrt{\sum_{ik}\left|A_{ik}\right|^{2}}\label{eq:HS-Norm-1}\end{equation}
is an equivalent definition. In that equation, $A_{ik}=\left\langle u_{i},u_{k}\right\rangle $
are the matrix elements in an arbitrary orthonormal basis $u_{i}$.
Operators with $\left\Vert \mathbf{A}\right\Vert _{2}<\infty$ are
called \emph{Hilbert-Schmidt operators}. The Hilbert-Schmidt norm
can also be written as\begin{equation}
\left\Vert \mathbf{A}\right\Vert _{2}=\sqrt{\sum_{i=1}^{\infty}\lambda_{i}^{2}}\,,\label{eq:HS-Norm-2}\end{equation}
where the $\lambda_{i}^{2}$ are the eigenvalues of $\mathbf{A}^{\dagger}\mathbf{A}$.
In case that $\mathbf{A}$ is self-adjoint, the real numbers $\lambda_{i}$
are just the eigenvalues of $\mathbf{A}$.
\item The \emph{trace norm }$\left\Vert \mathbf{A}\right\Vert _{1}$ for
a self-adjoint operator is given by \begin{equation}
\left\Vert \mathbf{A}\right\Vert _{1}=\sum_{i=1}^{\infty}\left|\lambda_{i}\right|\,,\label{eq:trace-norm}\end{equation}
where $\lambda_{i}$ are the eigenvalues of $\mathbf{A}$ and a purely
discrete spectrum of $\mathbf{A}$ has been assumed. Operators with
$\left\Vert \mathbf{A}\right\Vert _{1}<\infty$ are called \emph{trace
class operators}.
\end{enumerate}
In our paper, we make use of some properties of \emph{compact} operators.
For the exact mathematical definition of compactness, we refer to
\cite{Reed-Simon-1}. For our purposes, it is sufficient to recall
that every Hilbert-Schmidt operator is compact (see Theorem VI.22
in \cite{Reed-Simon-1}).

From the above definitions between norms, the relation\begin{equation}
\left\Vert \mathbf{A}\right\Vert _{2}^{2}=\left\Vert \mathbf{A}^{\dagger}\mathbf{A}\right\Vert _{1}\le\left\Vert \mathbf{A}\right\Vert _{1}^{2}\label{eq:Norm_HS_vs_trace}\end{equation}
follows immediately. Furthermore, there is the following hierarchical
relation holds between the norms defined above (\cite{Reed-Simon-1},
Theorem VI.22):\begin{equation}
\left\Vert \mathbf{A}\right\Vert \le\left\Vert \mathbf{A}\right\Vert _{2}\le\left\Vert \mathbf{A}\right\Vert _{1}\label{eq:normhierarchie}\end{equation}
The following inequalities also hold \begin{equation}
\left\Vert \mathbf{CAB}\right\Vert _{1}\le\left\Vert \mathbf{C}\right\Vert \left\Vert \mathbf{A}\right\Vert _{1}\left\Vert \mathbf{B}\right\Vert \,,\label{eq:CAB_1}\end{equation}
\begin{equation}
\left\Vert \mathbf{CAB}\right\Vert _{2}\le\left\Vert \mathbf{C}\right\Vert \left\Vert \mathbf{A}\right\Vert _{2}\left\Vert \mathbf{B}\right\Vert \,,\label{eq:CAB_2}\end{equation}
provided that all the norms exist.

\section{Proof of Theorem 2}

In this appendix, we proof the bound (\ref{eq:truncation_error-1})
for the approximation error of the eigenvalues as stated in Theorem
2. This bound is based on bounds for approximation errors for operators.

First of all, we shall proof that the operator $\tilde{\mathbf{R}}$
with matrix elements $\tilde{R}_{mn}$ given by Equation (\ref{eq:R_mn_tilde_FK})
is bounded.

\paragraph{Lemma 1:}

Under the assumption that $\mathcal{S}\left(\alpha\right)$ is a piecewise
continuous, bounded function over the interval $\left[-\pi,\pi\right]$,
the operator $\tilde{\mathbf{R}}$ with matrix elements $\tilde{R}_{mn}$
given by Equation (\ref{eq:R_mn_tilde_FK}) is bounded with operator
norm given by\[
\bigl\Vert\tilde{\mathbf{R}}\bigr\Vert=2\pi\max_{\alpha\in\left[-\pi,\pi\right]}\mathcal{S}\left(\alpha\right)\triangleq\rho_{max}\]

\paragraph{Proof:}

Since we have assumed that $\mathcal{S}\left(\alpha\right)$ is a
piecewise continuous, bounded function over the interval $\left[-\pi,\pi\right]$,
it defines a bounded multiplication operator \[
\mathbf{S}:\, X\left(\alpha\right)\mapsto Y\left(\alpha\right)=\mathcal{S}\left(\alpha\right)X\left(\alpha\right)\]
 on $\mathcal{L}^{2}\left(\left[-\pi,\pi\right],\frac{\mathrm{d}\alpha}{2\pi}\right)$.
Its operator norm is given by \[
\left\Vert \mathbf{S}\right\Vert =\max_{\alpha\in\left[-\pi,\pi\right]}\mathcal{S}\left(\alpha\right)\,.\]
The Hilbert space $\mathcal{L}^{2}\left(\left[-\pi,\pi\right],\frac{\mathrm{d}\alpha}{2\pi}\right)$
is isomorphic to the Hilbert space $\ell^{2}\left(\mathbb{Z}\right)$
of square summable sequences with indices running over $\mathbb{Z}$.
The vectors $X\left(\alpha\right),\, Y\left(\alpha\right)\in\mathcal{L}^{2}\left(\left[-\pi,\pi\right],\frac{\mathrm{d}\alpha}{2\pi}\right)$
correspond to the sequences $\left\{ x_{n}\right\} _{n\in\mathbb{Z}},\,\left\{ y_{n}\right\} _{n\in\mathbb{Z}}\in\ell^{2}\left(\mathbb{Z}\right)$
of their Fourier coefficients. The action of the operator $\mathbf{S}$
in this discrete representation is given by the convolution\[
y_{n}=\sum_{m=-\infty}^{\infty}\hat{s}_{m-n}x_{n}\,,\]
where $\hat{s}_{n}$ are the Fourier coefficients of $\mathcal{S}\left(\alpha\right)$.
Comparing with Equation (\ref{eq:R_mn_tilde_FK}) we see that $\tilde{\mathbf{R}}=2\pi\mathbf{S}$
holds in this represention. Since the operator norm does not depend
on the representation, we conclude that $\bigl\Vert\tilde{\mathbf{R}}\bigr\Vert=2\pi\left\Vert \mathbf{S}\right\Vert $,
which completes the proof. $\blacksquare$

For two compact and self-adjoint operators, the difference of the
corresponding eigenvalues is bounded by the operator norm of their
difference. We cite the following fact from the textbook of Riesz
and Nagy \cite{Riesz-Nagy}, § 95:

\paragraph{Lemma 2:}

Let $\mathbf{A}$ and \textbf{$\mathbf{B}$} be self-adjoint compact
operators and define \[
\mathbf{C=A+B}\,.\]
Assume that the eigenvalues $\lambda_{i}\left(\mathbf{A}\right)$
of $\mathbf{A}$ and $\lambda_{i}\left(\mathbf{C}\right)$ of $\mathbf{C}$
are ordered according to (\ref{eq:eigenvalue_ordering}). Then their
difference is bounded by \[
\left|\lambda_{i}\left(\mathbf{C}\right)-\lambda_{i}\left(\mathbf{A}\right)\right|\le\left\Vert \mathbf{B}\right\Vert \,\]
for all $i$.

\paragraph{Corolarry to Lemma 2:}

Setting $\mathbf{C}=\mathbf{R}$, $\mathbf{A}=\mathbf{R}_{N}$, and
$\mathbf{B}=\mathbf{R}-\mathbf{R}_{N}$ leads to the inequality\begin{equation}
\left|\lambda_{i}\left(\mathbf{R}\right)-\lambda_{i}\left(\mathbf{R}_{N}\right)\right|\le\left\Vert \mathbf{R}-\mathbf{R}_{N}\right\Vert \label{eq:Eigenwert-Abschaetzung}\end{equation}

The goal now is to show that $\left\Vert \mathbf{R}-\mathbf{R}_{N}\right\Vert \to0$
holds for $\ensuremath{N\to\infty}$. 

In the following, we assume that $\mathcal{A}\subset\mathcal{D}$,
where $\mathcal{D}$ is a disk of radius $r_{1}$. We define $N_{D}=\left\lceil \mathrm{e}\pi r_{1}\right\rceil $
and assume that $N\ge N_{D}$.

The transformation matrix $\mathbf{T}$ is defined by its elements
$T_{mn}=\left\langle u_{m},v_{n}\right\rangle $. The Gram matrix
$\mathbf{G}=\mathbf{T}^{\dagger}\mathbf{T}$ has matrix elements $G_{mn}=\left\langle v_{m},v_{n}\right\rangle $.
The squared Hilbert-Schmidt norm of $\mathbf{T}$ is given by\begin{equation}
\left\Vert \mathbf{T}\right\Vert _{2}^{2}=\left\Vert \mathbf{T}^{\dagger}\mathbf{T}\right\Vert _{1}=\left\Vert \mathbf{G}\right\Vert _{1}=\sum_{n=-\infty}^{\infty}\left\langle v_{n},v_{n}\right\rangle \,.\label{eq:T_2^2}\end{equation}
Using Equation (\ref{eq:Def_v_n(x)}) we may write\begin{eqnarray*}
\left\langle v_{n},v_{n}\right\rangle  & = & \int_{\mathcal{A}}\left|v_{n}\left(\mathbf{x}\right)\right|^{2}\mathrm{d}\mu\left(\mathbf{x}\right)\,.\\
 & = & \int_{\mathcal{A}}\mathrm{J}_{n}^{2}\left(2\pi\left|\mathbf{x}\right|\right)\mathrm{d}\mu\left(\mathbf{x}\right)\end{eqnarray*}
The integral is over the array $\mathcal{A}$ with $\mathrm{d}\mu\left(\mathbf{x}\right)$
given by Equation (\ref{eq:d_mu_curve}) or (\ref{eq:d_mu_area})
for the respective cases. Inserting the above expression into Equation
(\ref{eq:T_2^2}) leads to \begin{eqnarray*}
\left\Vert \mathbf{T}\right\Vert _{2}^{2} & = & \int_{\mathcal{A}}\sum_{n=-\infty}^{\infty}\mathrm{J}_{n}^{2}\left(2\pi\left|\mathbf{x}\right|\right)\mathrm{d}\mu\left(\mathbf{x}\right)\,.\end{eqnarray*}
The properties \[
\sum_{n=-\infty}^{\infty}\mathrm{J}_{n}^{2}\left(x\right)=1\]
and\foreignlanguage{english}{\[
\int_{\mathcal{A}}\mathrm{d}\mu\left(\mathbf{x}\right)=1\]
}then yield\begin{equation}
\left\Vert \mathbf{T}\right\Vert _{2}^{2}=1\,.\label{eq:T_2^2-1}\end{equation}

We now define a \emph{truncated transformation matrix} $\mathbf{T}_{N}$
that has the same matrix elements as $\mathbf{T}$ inside the \emph{middle
square} defined by $\left|m\right|\le N$ and $\left|n\right|\le N$.
All other matrix elements are set to zero. The \emph{remainder matrix}
$\mathbf{T}-\mathbf{T}_{N}$ has only non-vanishing matrix elements
outside that middle square. Its squared Hilbert-Schmidt norm is given
by\begin{eqnarray*}
\left\Vert \mathbf{T}-\mathbf{T}_{N}\right\Vert _{2}^{2} & = & \mathrm{tr}\left[\left(\mathbf{T}-\mathbf{T}_{N}\right)^{\dagger}\left(\mathbf{T}-\mathbf{T}_{N}\right)\right]\\
 & = & \mathrm{tr}\left(\mathbf{T}^{\dagger}\mathbf{T}+\mathbf{T}_{N}^{\dagger}\mathbf{T}_{N}-\mathbf{T}^{\dagger}\mathbf{T}_{N}-\mathbf{T}_{N}^{\dagger}\mathbf{T}\right)\\
 & = & \mathrm{tr}\left(\mathbf{T}^{\dagger}\mathbf{T}\right)+\mathrm{tr}\left(\mathbf{T}_{N}^{\dagger}\mathbf{T}_{N}\right)-\mathrm{tr}\left(\mathbf{T}^{\dagger}\mathbf{T}_{N}\right)-\mathrm{tr}\left(\mathbf{T}_{N}^{\dagger}\mathbf{T}\right)\\
 & = & \mathrm{tr}\left(\mathbf{T}^{\dagger}\mathbf{T}\right)-\mathrm{tr}\left(\mathbf{T}_{N}^{\dagger}\mathbf{T}_{N}\right)\\
 & = & \mathrm{tr}\left(\mathbf{T}^{\dagger}\mathbf{T}-\mathbf{T}_{N}^{\dagger}\mathbf{T}_{N}\right)\end{eqnarray*}
Here we have used the fact that $\mathbf{T}^{\dagger}\mathbf{T}_{N}=\mathbf{T}_{N}^{\dagger}\mathbf{T}_{N}$
and thus $\mathrm{tr}\left(\mathbf{T}^{\dagger}\mathbf{T}_{N}\right)=\mathrm{tr}\left(\mathbf{T}_{N}^{\dagger}\mathbf{T}_{N}\right)$.
The truncated Gram matrix $\mathbf{G}_{N}=\mathbf{T}_{N}^{\dagger}\mathbf{T}_{N}$
has the same matrix elements as $\mathbf{G}$ inside the inner square
and zero elements outside. Thus, the expression derived above can
be written as \begin{eqnarray*}
\left\Vert \mathbf{T}-\mathbf{T}_{N}\right\Vert _{2}^{2} & = & \mathrm{tr}\left(\mathbf{G}-\mathbf{G}_{N}\right)\\
 & = & \sum_{\left|n\right|>N}\left\langle v_{n},v_{n}\right\rangle \\
 & = & \int_{\mathcal{A}}\sum_{\left|n\right|>N}\mathrm{J}_{n}^{2}\left(2\pi\left|\mathbf{x}\right|\right)\mathrm{d}\mu\left(\mathbf{x}\right)\end{eqnarray*}
 From Theorem 1 we get the bound\[
\left\Vert \mathbf{T}-\mathbf{T}_{N}\right\Vert _{2}^{2}\le0.01\,\exp\left[2\left(N_{D}-N\right)\right]\,.\]

Taking the square root yields\begin{equation}
\left\Vert \mathbf{T}-\mathbf{T}_{N}\right\Vert _{2}\le0.1\,\exp\left(N_{D}-N\right)\,.\label{eq:T-T_N}\end{equation}

From \[
\left\Vert \mathbf{T}_{N}\right\Vert _{2}^{2}=\sum_{\left|n\right|\le N}\left\langle v_{n},v_{n}\right\rangle \]
and Equations (\ref{eq:T_2^2}) and (\ref{eq:T_2^2-1}) we conclude
that \begin{equation}
\left\Vert \mathbf{T}_{N}\right\Vert _{2}\le\left\Vert \mathbf{T}\right\Vert _{2}=1\,\label{eq:T_N_le_T}\end{equation}
holds. We define the truncated correlation matrix \begin{equation}
\mathbf{R}_{N}=\mathbf{T}_{N}\tilde{\mathbf{R}}\mathbf{T}_{N}^{\dagger}\,.\label{eq:def-R_N}\end{equation}
The Hilbert-Schmidt norm approximation error relative to $\mathbf{R}=\mathbf{T}\tilde{\mathbf{R}}\mathbf{T}^{\dagger}$
can now be estimated as follows:\begin{eqnarray*}
\left\Vert \mathbf{R}-\mathbf{R}_{N}\right\Vert _{2} & = & \left\Vert \mathbf{T}\tilde{\mathbf{R}}\mathbf{T}^{\dagger}-\mathbf{T}_{N}\tilde{\mathbf{R}}\mathbf{T}_{N}^{\dagger}\right\Vert _{2}\\
 & = & \left\Vert \mathbf{T}\tilde{\mathbf{R}}\left(\mathbf{T}-\mathbf{T}_{N}\right)^{\dagger}+\left(\mathbf{T}-\mathbf{T}_{N}\right)\tilde{\mathbf{R}}\mathbf{T}_{N}^{\dagger}\right\Vert _{2}\\
 & \le & \left\Vert \mathbf{T}\tilde{\mathbf{R}}\left(\mathbf{T}-\mathbf{T}_{N}\right)^{\dagger}\right\Vert _{2}+\left\Vert \left(\mathbf{T}-\mathbf{T}_{N}\right)\tilde{\mathbf{R}}\mathbf{T}_{N}^{\dagger}\right\Vert _{2}\\
 & \le & \left\Vert \mathbf{T}\right\Vert _{2}\bigl\Vert\tilde{\mathbf{R}}\bigr\Vert\left\Vert \mathbf{T}-\mathbf{T}_{N}\right\Vert _{2}+\left\Vert \mathbf{T}-\mathbf{T}_{N}\right\Vert _{2}\bigl\Vert\tilde{\mathbf{R}}\bigr\Vert\left\Vert \mathbf{T}_{N}\right\Vert _{2}\\
 & \le & 2\bigl\Vert\tilde{\mathbf{R}}\bigr\Vert\left\Vert \mathbf{T}-\mathbf{T}_{N}\right\Vert _{2}\end{eqnarray*}
From the bound (\ref{eq:T-T_N}) we conclude \begin{equation}
\left\Vert \mathbf{R}-\mathbf{R}_{N}\right\Vert _{2}\le0.2\cdot\bigl\Vert\tilde{\mathbf{R}}\bigr\Vert\exp\left(N_{D}-N\right)\,.\label{eq:R-R_N}\end{equation}
The norm hierarchy (\ref{eq:normhierarchie}) yields\begin{equation}
\left\Vert \mathbf{R}-\mathbf{R}_{N}\right\Vert \le0.2\cdot\bigl\Vert\tilde{\mathbf{R}}\bigr\Vert\exp\left(N_{D}-N\right)\,.\label{eq:R-R_N-1}\end{equation}
From the Corollary to Lemma 2 we obtain\[
\left|\lambda_{i}\left(\mathbf{R}\right)-\lambda_{i}\left(\mathbf{R}_{N}\right)\right|\le0.2\cdot\left\Vert \tilde{\mathbf{R}}\right\Vert \exp\left(N_{D}-N\right)\,.\]
Lemma 1 then leads to Theorem 2.

\section{Proof of Theorem 3}

In this appendix, we prove the bound (\ref{eq: HS-Norm-Schranke})
for the approximation error of the squared Hilbert Schmidt norms as
stated in Theorem 3. From $\mathbf{R}=\mathbf{T}\tilde{\mathbf{R}}\mathbf{T}^{\dagger}$
and $\mathbf{R}_{N}=\mathbf{T}_{N}\tilde{\mathbf{R}}\mathbf{T}_{N}^{\dagger}$
and Inequality (\ref{eq:T_N_le_T}) we conclude\[
\left\Vert \mathbf{R}\right\Vert _{2}\le\bigl\Vert\tilde{\mathbf{R}}\bigr\Vert\mbox{ and }\left\Vert \mathbf{R}_{N}\right\Vert _{2}\le\bigl\Vert\tilde{\mathbf{R}}\bigr\Vert\,.\]
Then \begin{eqnarray*}
\left|\left\Vert \mathbf{R}\right\Vert _{2}^{2}-\left\Vert \mathbf{R}_{N}\right\Vert _{2}^{2}\right| & = & \Bigl|\left\Vert \mathbf{R}\right\Vert _{2}-\left\Vert \mathbf{R}_{N}\right\Vert _{2}\Bigr|\left(\left\Vert \mathbf{R}\right\Vert _{2}+\left\Vert \mathbf{R}_{N}\right\Vert _{2}\right)\\
 & \le & 2\bigl\Vert\tilde{\mathbf{R}}\bigr\Vert\Bigl|\left\Vert \mathbf{R}\right\Vert _{2}-\left\Vert \mathbf{R}_{N}\right\Vert _{2}\Bigr|\end{eqnarray*}
The bound (\ref{eq:R-R_N}) yields\[
\left|\left\Vert \mathbf{R}\right\Vert _{2}^{2}-\left\Vert \mathbf{R}_{N}\right\Vert _{2}^{2}\right|\le0.4\bigl\Vert\tilde{\mathbf{R}}\bigr\Vert^{2}\exp\left(N_{D}-N\right)\,.\]
With Lemma 1 we then obtain Theorem 3.

\bibliographystyle{ieeetr}
\bibliography{/home/schulze/Papers/DABDIVDOC/Diversity_Spectra/div_spectra}

\begin{thebibliography}{10}

\bibitem{Proakis}
J.~G. Proakis and M.~Salehi, {\em Digital Communications}.
\newblock McGraw-Hill, fifth~ed., 2008.

\bibitem{BenedettoBiglieri}
S.~Benedetto and E.~Biglieri, {\em Principles of Digital Transmission With
  Wireless Applications}.
\newblock Kluwer, 1999.

\bibitem{Rappaport}
T.~S. Rappaport, {\em Wireless Communications: Principles and Practice}.
\newblock Prentice Hall, second~ed., 2001.

\bibitem{SchulzeLueders}
H.~Schulze and C.~L\"uders, {\em Theory and Practice of OFDM and CDMA -
  Wideband Wireless Communications}.
\newblock Wiley, 2005.

\bibitem{Foschini1998}
G.~Foschini and M.~Gans, ``On limits of wireless communications in a fading
  environment when using multiple antennas,'' {\em Wireless Personal
  Communications}, vol.~6, pp.~311--335, March 1998.

\bibitem{Telatar1999}
I.~E. Telatar, ``Capacity of multi-antenna {G}aussian channels,'' {\em European
  Transactions on Telecommunications}, vol.~10, pp.~585--595, December 1999.

\bibitem{KuehnBook}
V.~K\"uhn, {\em Wireless Communications over {MIMO} Channels: Applications to
  CDMA and Multiple Antenna Systems}.
\newblock Wiley, 2006.

\bibitem{Van_Trees}
H.~{Van Trees}, {\em Detection, estimation, and modulation theory, {Part I}}.
\newblock Wiley, 2001.

\bibitem{Davenport_Root}
W.~B. Davenport and W.~Root, {\em An Introduction to the theory of random
  signals and noise}.
\newblock McGraw-Hill, 1958.

\bibitem{Papoulis}
A.~Papoulis, {\em Probability, Random Variables, and Stochastic Processes}.
\newblock McGraw-Hill, 1991.

\bibitem{Schulze2010a}
H.~Schulze, ``Matched filter bounds for multi-antenna {OFDM} systems,'' in {\em
  Proc. 15 th Int. OFDM Workshop (Hamburg)}, 2010.

\bibitem{Ivrlac2003}
M.~T. Ivrla\v{c} and J.~A. Nossek, ``Quantifying diversity and correlation in
  {R}ayleigh fading {MIMO} communication systems,'' in {\em International
  Symposium on Signal Processing and Information Theory}, 2003.

\bibitem{Muharemovic2008}
T.~Muharemovic, A.~Sabharwal, and B.~Aazhang, ``Antenna packing in low-power
  systems: Communication limits and array design,'' {\em IEEE Trans. Inform.
  Theory}, vol.~IT-54(1), pp.~429--440, January 2008.

\bibitem{Lozano2003}
A.~Lozano, A.~M. Tulino, and S.~Verd{u}, ``Multiple-antenna capacity in the
  low-power regime,'' {\em IEEE Trans. Inform. Theory}, vol.~IT-49(10),
  pp.~2527--2544, October 2003.

\bibitem{Kennedy2007}
R.~A. Kennedy, P.~Sadeghi, T.~D. Abhayapala, and H.~M. Jones, ``Intrinsic
  limits of dimensionality and richness in random multipath fields,'' {\em IEEE
  Trans. on Signal Processing}, vol.~55 (6), pp.~2542--2556, June 2007.

\bibitem{ArfkenWeber}
G.~B. Arfken and H.~J. Weber, {\em Mathematical Methods for Physicists}.
\newblock Elsevier, 6th~ed., 2005.

\bibitem{Bello1963}
P.~A. Bello, ``Characterization of randomly time-variant linear channels,''
  {\em IEEE Trans. Commun. Syst.}, vol.~CS-11, pp.~360--393, Dec. 1963.

\bibitem{Fleury2000}
B.~H. Fleury, ``First- and second-order characterization of direction
  dispersion and space selectivity in the radio channel,'' {\em IEEE Trans.
  Inf. Theory}, vol.~46(6), pp.~2027--2047, September 2000.

\bibitem{Jakes}
W.~C. Jakes, {\em Microwave Mobile Communications}.
\newblock Wiley-Interscience, 1974.

\bibitem{Reed-Simon-1}
M.~Reed and B.~Simon, {\em Methods of modern mathematical physics, Vol. 1:
  Functional Analysis}.
\newblock Academic Press, 1980.

\bibitem{Courant_Hilbert-1}
R.~Courant and D.~Hilbert, {\em Methoden der {Mathematischen Physik}, Bd. 1}.
\newblock Springer-Verlag, 1968.

\bibitem{Riesz-Nagy}
F.~Riesz and B.~Sz.-Nagy, {\em Functional Analysis}.
\newblock Dover, 1990.

\bibitem{Horn-Johnson}
R.~A. Horn and C.~R. Johnson, {\em Matrix Ananlysis}.
\newblock Cambridge, 1985.

\bibitem{Ravindran}
P.~Ravindran, {\em Bayesian Analysis of Circular Data Using Wrapped
  Distributions}.
\newblock Dissertation, North Carolina State University, 2002.

\bibitem{AbramowitzStegun}
M.~Abramowitz and I.~Stegun, {\em Handbook of Mathematical Functions}.
\newblock Dover, 1964.

\end{thebibliography}

\end{document}